\pdfoutput=1
\documentclass[a4paper,11pt]{article}

\usepackage[left=2.5cm, right=2.5cm, top=3.0cm, bottom=3.0cm]{geometry}

\usepackage{amsmath,amssymb,amsfonts,mathrsfs}

\usepackage{todonotes}
\usepackage{tikz}
\usepackage{tikz-cd}
\usetikzlibrary{cd}
\usepackage[all]{xy}
\usetikzlibrary{shapes,arrows}
\usetikzlibrary{shapes}
\usetikzlibrary{plotmarks}
\usepackage{tikz-3dplot}
\usetikzlibrary{calc}
\usetikzlibrary{arrows.meta,positioning}

\usepackage{jheppub}
\usepackage{verbatim}
\usepackage{dsfont}

\def\cD{\mathcal{D}}

\def\cT{\mathcal{T}}

\graphicspath{ {./Figures/} }

\usepackage{hyperref}

\newcommand{\appref}[1]{\hyperref[#1]{Appendix~\ref*{#1}}}

\title{Half-Spacetime Gauging of 2-Group Symmetry in 3d}
\author[1]{Davide Bason,}
\author[2]{Wei Cui}
\author[3,4]{and Lorenzo Ruggeri}
\affiliation[1]{Yau Mathematical Sciences Center, Tsinghua University, Jingzhai, Haidian District, Beijing, 100084, China}
\affiliation[2]{Beijing Institute of Mathematical Sciences and Applications (BIMSA), Huairou District, Beijing
101408, China}
\affiliation[3]{Dipartimento di Matematica “Giuseppe Peano”, Universit\`a di Torino, Via Carlo Alberto 10,
10123 Torino, Italy}
\affiliation[4]{INFN, Sezione di Torino, Via Pietro Giuria 1, 10125 Torino, Italy}

\emailAdd{davidebason@mail.tsinghua.edu.cn}
\emailAdd{cwei@bimsa.cn}
\emailAdd{lorenzo.ruggeri@unito.it}

\abstract{We construct a class of non-invertible duality defects, in (2+1)d quantum field theories, arising from half-spacetime gauging of a 2-group symmetry. Starting from a parent theory with two discrete and Abelian 0-form symmetries and a prescribed mixed anomaly, we show that gauging one factor produces a theory with a 2-group symmetry, while gauging the other yields a theory with a non-invertible 0-form symmetry, whose fusion rules we derive explicitly. When the parent theory possesses three such symmetries with a cyclic anomaly structure, gauging different factors can produce mutually dual theories and the half-spacetime gauging of the 2-group is implemented by a non-invertible duality defect, whose fusion rules we obtain. We illustrate the construction with explicit examples, including a $U(1)\times U(1)\times U(1)$ gauge theory and a general class of product theories. We also include a self-contained pedagogical introduction to the cohomological tools employed throughout the article.}

\begin{document}

\maketitle{}

\flushbottom

\section{Introduction}
The modern description of a global symmetry involves specifying its conserved charges, or symmetry operators, rather than realizing it by Lagrangians and fields \cite{Gaiotto:2014kfa}. See \cite{Schafer-Nameki:2023jdn,Shao:2023gho,Kaidi:2026urc} for a selection of reviews on this vast and ever-growing topic. A $q$-form symmetry is implemented by a codimension-$(q+1)$ topological defect. If the fusion of two topological defects follows a group-like structure, the corresponding symmetry is invertible. However, as it has been known for many years in the context of 2d rational CFTs \cite{Verlinde:1988sn,Moore:1988qv,Moore:1989yh}, this does not need to be the case. In fact, the fusion of two defects can give rise to a linear combination of more than one defect, giving rise to \emph{non-invertible symmetries} \cite{Bhardwaj:2017xup,Chang:2018iay,Tachikawa:2017gyf,Kaidi:2021xfk,Choi:2021kmx,Choi:2022zal}. 

Another structure that naturally emerges from this framework is that of a higher group. When a theory contains several higher-form symmetries, their combined symmetry structure does not need to be a direct product; instead, the symmetries may be non-trivially intertwined, giving rise to a \emph{higher-group symmetry}. Here, we restrict to 2-groups formed by a 0-form and a 1-form symmetry. The higher-categorical structure was first introduced in \cite{Baez:2003yaq,Baez:2004in,Baez:2005qu,Schreiber:2008kcv,Etingof:2009yvg,Sharpe:2015mja}. In the context of physics, SPT phases with discrete 2-group symmetry were constructed in \cite{Kapustin:2013uxa}, with subsequent generalizations provided in\footnote{In the condensed matter literature, a 2-group global symmetry is referred to as an obstruction to symmetry fractionalization \cite{Barkeshli:2014cna,Barkeshli:2017rzd}.} \cite{Thorngren:2015gtw,Gaiotto:2017zba}. Further studies have explored 2-group symmetries for both discrete groups \cite{Bhardwaj:2016clt,Tachikawa:2017gyf,Delcamp:2018wlb} and continuous groups \cite{Cordova:2018cvg}, while a generic approach using the language of symmetry defects was developed in \cite{Benini:2018reh}. More recent works include \cite{Hsin:2020nts,Apruzzi:2021mlh,Lee:2021crt,Santilli:2024dyz,Arbalestrier:2025poq,Inamura:2026hjl,Villa:2026jmd} Our main goal is to study the interplay between higher group structures and non-invertible symmetries.

In this work, we study two different ways of constructing non-invertible symmetry defects. The first method, pioneered by \cite{Kaidi:2021xfk,Koide:2021zxj}, consists in considering a theory in which two symmetries have a specific type of mixed 't Hooft anomaly. Upon gauging one of the two symmetries, the symmetry defect associated with the remaining one is no longer gauge invariant, but instead transforms non-trivially under gauge transformations. To overcome this issue, one stacks the defect with a non-invertible TQFT canceling the gauge anomaly. The non-invertible fusion rules follow from the non-invertibility of the TQFT. In our construction, such defects arise when considering two 0-form discrete and Abelian symmetries $G_A^{(0)}\times G_B^{(0)}$, and an anomaly theory as in \eqref{eq.parentanomaly}. Gauging $G_B^{(0)}$ results into a 2-group \cite{Kapustin:2013uxa,Cordova:2018cvg,Benini:2018reh}. Instead, gauging $G_A^{(0)}$ gives rise, upon stacking with a suitable TQFT, to a non-invertible symmetry which, to the best our knowledge, has not been explicitly studied before. We also compute the fusion rules of the non-invertible defect.
 
Another particularly well-known class of non-invertible categorical symmetries is realized by \emph{non-invertible duality defects}. If a theory admits an anomaly-free $q$-form symmetry, it is possible to consider an interface implementing its gauging in half of spacetime. Whenever the resulting theory is isomorphic to the original theory, the interface gives rise to a non-invertible duality defect.\footnote{Whenever the non-invertible duality defect can be equivalently described starting from a theory with invertible symmetries and a mixed 't Hooft, as the class described above, the corresponding symmetry is named non-intrinsically non-invertible or group-theoretical \cite{Sun:2023xxv}. If this is not the case, the non-invertible symmetry is denoted intrinsic. It would be interesting to understand whether the duality defects we consider in this work are intrinsic or not.} Instances of non-invertible symmetries which can be obtained via this construction include Kramers-Wannier duality in the Ising model \cite{Verlinde:1988sn,Aasen:2016dop,Freed:2018cec} and $\mathcal{N}=4$ Super Yang-Mills \cite{Choi:2021kmx,Choi:2022zal}. The initial constructions focused on theories with a Pontryagin self-dual $q$-form symmetry. This fixes $q=d/2-1$, hence even dimensions, and the discrete group to be a product of $\mathbb{Z}_N$. Constructions of non-invertible duality defects in odd dimensions have been initiated in \cite{Choi:2024rjm,Cui:2024cav}, with a focus on three-dimensional theories. As in 3d no symmetry gives back itself upon gauging, the idea is to simultaneously gauge a 0-form and a 1-form symmetry, which exchange roles under gauging. 

In the present article we extend this construction by allowing a non-trivial 2-group structure between the 0-form and 1-form symmetries. In cases where the theories before and after gauging are isomorphic, half-spacetime gauging provides the first explicit duality defect arising from the gauging of a 2-group. For this to happen, the theory must contain three symmetry sectors: two combine into a 2-group symmetry, while the third is generated by the non-invertible symmetry defect obtained by stacking with a TQFT we introduced above. Upon gauging the 2-group, these sectors are reshuffled. The dual 2-group is formed by the original non-invertible 0-form symmetry, which becomes its 0-form component, together with the original 0-form symmetry of the gauged 2-group, which is mapped to its 1-form component. The remaining original 1-form symmetry is instead realized as a non-invertible 0-form symmetry. The duality defect implementing the gauging has non-invertible fusion rules, which we derive explicitly. Such theory can be obtained from a parent theory $\mathcal{T}$ with symmetry group $G_A^{(0)}\times G_B^{(0)}\times G_C^{(0)}$, and an anomaly theory as in \eqref{eq.4danomT}, upon gauging one of the three symmetry groups.

We also include in the appendices a pedagogical introduction to the cohomological tools used throughout, in order to make the article more self-contained and accessible to readers less familiar with the subject.
 
The remainder of this paper is organized as follows. In \autoref{sec.2} we introduce a prototypical parent theory $\mathcal{T}$ with two discrete 0-form symmetries and a prescribed mixed anomaly. We then analyze the consequences of gauging each symmetry independently: gauging $G_B^{(0)}$ gives rise to a 2-group structure, while gauging $G_A^{(0)}$ leads to a non-invertible symmetry. In \autoref{sec.3} we use these results to construct a duality defect via half-spacetime gauging of the 2-group and we derive the corresponding non-trivial fusion rules. In \autoref{sec.4} we illustrate the construction with explicit examples, including a $U(1)\times U(1)\times U(1)$ gauge theory and a general class of product theories. We conclude in \autoref{sec.5} with a summary of our results and a discussion of future directions. Technical details, conventions and supplementary computations are collected in the appendices.

\section{Parent Theory and its Descendants}\label{sec.2}
In this section, we describe a (2+1)d theory $\mathcal{T}$, possessing two Abelian discrete $0$-form symmetries $G_A^{(0)}\times G_{B}^{(0)}$, and with the following prescribed anomaly theory
\begin{equation}\label{eq.parentanomaly}
    S_\text{anomaly}=\int_{Y_4}B^{(1)} A^{(1)*}\beta\,.
\end{equation}
Here, $A^{(1)}$ and $B^{(1)}$ are background fields for $G_A^{(0)}\times G_{B}^{(0)}$. Moreover, $\beta\in H^3(K^1G_A^{(0)},\hat{G}_B^{(1)})$ is a Postnikov class, making $A^{(1)*}\beta$ a 3-cocycle,\footnote{To be more specific, this makes use of the dual interpretation of an element in the cohomology $A^{(1)}$ as a map from $Y_4$ to $K^1G_A^{(0)}$, which we use to pull back $\beta$ onto $Y_4$. Refer to \appref{aeilenbermaclane} for more details.} and $\hat{G}_B^{(1)}$ is the Pontryagin dual to $G_B^{(0)}$. The coupling is the Steenrod cup product discussed in \eqref{estenroodscup}-\eqref{eq.esteenrodpontr}. Throughout this article we largely omit writing explicitly cup products. Such theories have been originally considered in \cite{Kapustin:2013uxa,Tachikawa:2017gyf,Benini:2018reh}. Our first goal is to study the effect of gauging the discrete Abelian symmetries of the theory in the presence of the prescribed anomaly. We refer the reader to \appref{aconventions} for our conventions and to \appref{aintrocohomology} for an introduction to the tools used throughout this article.

The anomaly manifests as a non-trivial relative phase factor of the partition function computed in the presence of topologically equivalent defect networks \cite{Tachikawa:2017gyf}. More concretely, it is a phase obtained upon shifting the background gauge fields describing the defect networks by exact cochains
\begin{equation}
    A^{(1)}\to A^{(1)}+\delta\lambda_A^{(1)}, \qquad B^{(1)}\to B^{(1)}+\delta\lambda_B^{(1)}.
\end{equation}
Such anomaly is captured by a closed cochain in one dimension higher. Explicitly, this means that
\begin{align}\label{eq.anomalyofourtheory}
    Z[A^{(1)}+\delta\lambda_A^{(1)},B^{(1)}+\delta\lambda_B^{(1)}]=e^{ i\int_{X_3}\lambda_B^{(0)} (A^{(1)*}\beta+\delta \zeta^{(2)}(A^{(1)},\lambda_A^{(0)}))-B^{(1)}\zeta^{(2)}(A^{(1)},\lambda_A^{(0)})}Z[A^{(1)},B^{(1)}]\,.
\end{align}
Here, we used that
\begin{align}\label{eq.descendant}
    (A^{(1)}+\delta\lambda_A^{(0)})^*\beta-A^{(1)*}\beta=\delta\zeta^{(2)}(A^{(1)},\lambda_A^{(0)})\,,
\end{align}
for some 2-cochain $\zeta^{(2)}$ specified by $\beta$ via descent relations \cite{Benini:2018reh, Kapustin:2013uxa}. Notice that shifting $\zeta^{(2)}$ by exact 2-cochains leaves \eqref{eq.anomalyofourtheory} invariant.
For this reasons, we refer interchangeably to this change in representatives and invariance thereof, as gauge transformation and gauge invariance respectively.

We proceed by analyzing gauging $G_B^{(0)}$ and $G_A^{(0)}$ separately. While gauging $G_B^{(0)}$ results into a theory with a 2-group symmetry, as described extensively in \cite{Kapustin:2013uxa,Benini:2018reh}, the study of the latter case leads to a non-invertible symmetry, which is a novel result of this article. However, as in the next section we study the gauging of a 2-group, we present also the former case.

\subsection{2-Group Symmetry}\label{sec.2.1}
In \appref{app.gaugingnonan}, we present a detailed account on gauging a non-anomalous symmetry. In this subsection, and in the subsequent one, we follow the presentation found there, to which we refer for more details. Upon gauging $G_B^{(0)}$, the resulting theory possesses a 2-group structure $\mathbb{G}=(G_{A}^{(0)},\hat{G}_{B}^{(1)},1,[\beta])$.\footnote{More generic 2-groups $\mathbb{G}=(G_{A}^{(0)},\hat{G}_{B}^{(1)},\rho,[\beta])$ also include a group homomorphism $\rho:G_A^{(0)}\to\text{Aut}(\hat{G}_{B}^{(1)})$ acting on an operator $\mathfrak{b}\in\hat{G}_{B}^{(1)}$ such that, when it pierces a symmetry defect $\mathfrak{a}\in G_{A}^{(0)}$, it emerges as a new 1-form charge $\rho_\mathfrak{a}\mathfrak{b}$. In the construction we consider in this work, obtained upon gauging a 0-form symmetry in $\mathcal{T}$, $\rho$ is set to be the identity map.} The action of the class $[\beta]$ is as follows: when three defects $\mathfrak{a}_1,\mathfrak{a}_2,\mathfrak{a}_3\in G_A^{(0)}$ meet at a codimension-3 junction, the 1-form symmetry $\hat{G}_{B}^{(1)}$ is also present and it is controlled by $\beta$. These rules are encoded in the \textit{twisted cocycle condition} of the backgrounds gauge fields \cite{Kapustin:2013uxa}
\begin{equation}\label{e2groupbg}
    \delta\hat{B}^{(2)}=A^{(1)*}\beta\,.
\end{equation}
We refer to \cite{Kapustin:2013uxa,Benini:2018reh} for a detailed description of background fields for 2-groups.

\paragraph{Gauging $G_B^{(0)}$.}
We start by choosing a complete collection of independent representatives $b\in H^1_B=H^1(X_3,G^{(0)}_B)$ and define the partition function of gauged theory $\mathcal{T}_B\equiv\mathcal{T}/G_B^{(0)}$ as 
\begin{align}\label{eq.tentativeq.gauging1}
    Z_b^B[A^{(1)},\hat{B}^{(2)}]=\frac{1}{|H_B^0|}\underset{b\in H^1_B}{\sum}e^{i\int_{X_3} b^{(1)}\hat{B}^{(2)}}Z[A^{(1)},b^{(1)}]\,.
\end{align}
As in \eqref{edefectrepresentation}, the nature of the cochain $\hat{B}^{(2)}$ is fixed by requiring gauge invariance $Z_b^B=Z_{b+\delta \lambda_b}^B$, which guarantees that the all physical observables are independent of the $G_B^{(0)}$-action. 

Under a gauge transformation the partition function transforms as follows
\begin{align}
    \begin{split}
    Z_{b+\delta\lambda_b}^B[A^{(1)},\hat{B}^{(2)}]&=\frac{1}{|H_B^0|}\underset{b\in H^1_B}{\sum}e^{ i\int_{X_3}\left(b^{(1)}+\delta\lambda^{(0)}_b\right)\hat{B}^{(2)}}Z[A^{(1)},b^{(1)}+\delta\lambda^{(0)}_b]\\
    &=\frac{1}{|H_B^0|}\underset{b\in H^1_B}{\sum}e^{i\int_{X_3} b^{(1)}\hat{B}^{(2)}-\lambda^{(0)}_b\delta\hat{B}^{(2)}}Z[A^{(1)},b^{(1)}+\delta\lambda_b^{(0)}]\\
    &=\frac{1}{|H_B^0|}\underset{b\in H^1_B}{\sum}e^{i\int_{X_3}b^{(1)}\hat{B}^{(2)}}e^{i\int_{X_3}\lambda_b^{(0)} (A^{(1)*}\beta-\delta\hat{B}^{(2)})}Z[A^{(1)},b^{(1)}]\,.
    \end{split}
\end{align}
In the second line we have used integration by parts while in the third line we used \eqref{eq.anomalyofourtheory}. As we mentioned above, we find that the gauge-invariance of $Z_b^B$ enforces the twisted cocycle condition \eqref{e2groupbg}. Moreover, as we have shown that the partition function is independent on $b$, we denote it simply by $Z^B$.

Due to the descent relation \eqref{eq.descendant}, $Z^B$ is covariant under 
\begin{equation}\begin{split}\label{eq.2groupbgtransf}
    &A^{(1)}\rightarrow A^{(1)}+\delta\lambda_{A}^{(0)}\,,\\
    &\hat{B}^{(2)}\rightarrow\hat{B}^{(2)}+\delta\lambda_{\hat{B}}^{(1)}+\zeta^{(2)}(A^{(1)},\lambda_A^{(0)})\,,
\end{split}\end{equation}
which, as it can be checked explicitly, implies the following
\begin{align}\label{e2groupnoanomaly}
    Z^B[A^{(1)}+\delta\lambda_A^{(0)},\hat{B}^{(2)}+\delta\lambda^{(1)}_{\hat{B}}+\zeta^{(2)}(A^{(1)},\lambda_A^{(0)})]=Z^B[A^{(1)},\hat{B}^{(2)}]\,.
\end{align}
As we argued below \eqref{eq.descendant}, $\zeta^{(2)}(A^{(1)},\lambda_A^{(0)})$ is defined modulo exact 2-cocycles; here this ambiguity is reflected in the fact that any such shift can be absorbed into $\delta\lambda^{(1)}_B$.

\paragraph{Gauging $\hat{G}_B^{(1)}$.}
As just discussed, the theory $\mathcal{T}/G_B^{(0)}$ has a 2-group structure. Its gauging is studied in the next section, here we consider the gauging of its 1-form symmetry $\hat{G}_B^{(1)}$.\footnote{This gauging, however, depends explicitly on $A^{(1)}$, namely on the $G_A^{(0)}$ defect-network we might keep in the vacuum.} Our goal is to show that this operation produces the original theory with the anomaly in \eqref{eq.anomalyofourtheory}. Repeating similar steps as above we write
\begin{align}\label{eq.gauginggbhat}
    Z_{\hat{b}}^{B,\hat{B}}[A^{(1)},\hat{\hat{B}}^{(1)}]=\frac{|H^0_{\hat{B}}|}{|H^1_{\hat{B}}|}\underset{\hat{b}\in H_{\delta\hat{b}=A^*\beta}^2}{\sum}e^{i\int_{X_3}\hat{b}^{(2)}\hat{\hat{B}}^{(1)}}Z^B[A^{(1)},\hat{b}^{(2)}],
\end{align}
for some $\hat{\hat{B}}^{(1)}\in C^1_B$, whose constraints we discussed below. We have denoted as
\begin{align}
\begin{split}
    &Z^2_{\delta \hat{b}=A^*\beta}=\{\forall\,\hat{b}^{(2)}\in C^2_{\hat{B}}\,|\,\delta\hat{b}^{(2)}=A^{(1)*}\beta\}\,,\\
    &H^2_{\delta \hat{b}=A^*\beta}=\frac{Z^2_{\delta \hat{b}=A^*\beta}}{B^2_{\hat{B}}}\,,
\end{split}
\end{align}
the twisted cohomology, so the sum runs only over these representatives. Imposing gauge invariance fixes $\delta\hat{\hat{B}}^{(1)}=0$. Due to \eqref{e2groupnoanomaly}, we have
\begin{align}
    Z_{\hat{b}}^{B,\hat{B}}[A^{(1)},\hat{b}^{(2)}+\delta\lambda_b^{(1)}]=Z_{\hat{b}}^{B,\hat{B}}[A^{(1)},\hat{b}^{(2)}]\,,
\end{align}
so we can exchange the sum over $\hat{b}\in H_{\delta\hat{b}=A^*\beta}^2$ with the one over $\hat{b}\in Z_{\delta\hat{b}=A^*\beta}^2$, provided that we normalize by the dimension of the space of exact 2-cocycles $|B_{\hat{B}}^2|$
\begin{align}\label{ealmostgauging2group}
    Z_{\hat{b}}^{B,\hat{B}}[A^{(1)},\hat{\hat{B}}^{(1)}]=\frac{|H^0_{\hat{B}}|}{|H^1_{\hat{B}}||B_{\hat{B}}^2|}\underset{\hat{b}\in Z_{\delta\hat{b}=A^*\beta}^2}{\sum}e^{i\int_{X_3}\hat{b}^{(2)}\hat{\hat{B}}^{(1)}}Z^B[A^{(1)},\hat{b}^{(2)}]\,.
\end{align}
Using \eqref{esomerelations}, and adding a Lagrange multiplier $\chi^{(0)}$ to fix the twisted cocycle conditions, we get
\begin{align}\label{eq.gauginggbhatunconstr}
    Z^{B,\hat{B}}[A^{(1)},\hat{\hat{B}}^{(1)}]=\frac{|H^0_{\hat{B}}||Z^1_{\hat{B}}|}{|H^1_{\hat{B}}||C^1_{\hat{B}}||C^0_B|}\underset{\hat{b}\in C^2_{\hat{B}},\chi\in C^0_B}{\sum}e^{i\int_{X_3}\chi^{(0)}(\delta\hat{b}^{(2)}-A^{(1)*}\beta)+\hat{b}^{(2)}\hat{\hat{B}}^{(1)}}Z^B[A^{(1)},\hat{b}^{(2)}]\,,
\end{align}
where we removed the $\hat{b}$ suffix from $Z_{\hat{b}}^{B,\hat{B}}$ as gauge invariance is now manifest. 

Sending $\hat{\hat{B}}\rightarrow -B$, we notice that, under a gauge transformation of the background gauge field, we recover the same anomaly as in \eqref{eq.anomalyofourtheory}. This is the analogue of gauging the quantum symmetry in the non-anomalous theory, see e.g. \eqref{edoubleq.gaugingnonan}. Indeed, letting $\chi_B$ be the Euler counterterm \eqref{eeulercounterterm}, one can show that
\begin{align}\label{eq.gaugingbackB}
    Z^{B,\hat{B}}[A^{(1)},\hat{\hat{B}}^{(1)}]=\chi^{-1}_B Z[A^{(1)},-\hat{\hat{B}}^{(1)}]\,,
\end{align}
which is what we expect in general upon gauging a symmetry and then its quantum symmetry. We have then shown that gauging twice the symmetry $G_B^{(0)}$ gives back the parent theory $\mathcal{T}$.

\subsection{Non-Invertible Symmetry}\label{sec.2.2}
We now present the novel case, that obtained by gauging $G_{A}^{(0)}$. In the resulting theory $\mathcal{T}_A\equiv\mathcal{T}/G_A^{(0)}$, the symmetry $G_B^{(0)}$ becomes non-invertible. Note that, a non-invertible defect network cannot be encoded in cocycles because of their composition rules. For future convenience, however, we use a cocycle $B$ associated to a single element of $G_B^{(0)}$ and a single 2-cycle, so as to uniquely and concisely specify an $\mathcal{N}$-defect. A generic defect network is described simply by a product of all these possible defects.

\paragraph{Gauging $G_{A}^{(0)}$.}
A naive attempt to gauge $G_A^{(0)}$ results in the following partition function
\begin{align}\label{egaugingnoninv}
    Z_a^A[\hat{A}^{(2)},B^{(1)}]=\frac{1}{|H^0_A|}\underset{a\in H^1_A}{\sum}e^{i\int_{X_3}a^{(1)}\hat{A}^{(2)}}Z[a^{(1)},B^{(1)}]\,.
\end{align}
Under a gauge transformation $Z_a^A$ transforms as
\begin{align}\label{etentativegaugeandshift}
    Z_{a+\delta\lambda_a}^A[\hat{A}^{(2)},B^{(1)}]=\frac{1}{|H^0_A|}\underset{a\in H^1_A}{\sum}e^{i\int_{X_3}a^{(1)}\hat{A}^{(2)}-\lambda^{(0)}_a\delta\hat{A}^{(2)}-B^{(1)} \zeta^{(2)}(a^{(1)},\lambda^{(0)}_a)}Z[a^{(1)},B^{(1)}]\,.
\end{align}
Imposing $\delta\hat{A}^{(2)}=0$ makes sure that the second term in the exponential vanishes. Furthermore $\hat{A}^{(2)}\rightarrow \hat{A}^{(2)}+\delta\lambda^{(1)}_{\hat{A}}$ does not change the partition function, and thus we obtain the expected $\hat{G}_A^{(1)}$ quantum symmetry. 

However, no background shift of $B^{(1)}$ and $\hat{A}^{(2)}$ in \eqref{etentativegaugeandshift} can make the third contribution in the exponential vanish. To fix this issue, we first notice that $B^{(1)}\zeta^{(2)}$ is supported on the $D_B$-network of $G_B^{(0)}$-transformations (see discussion in \appref{app.eresfourpoinc}). As pointed out in \cite{Choi:2021kmx, Kaidi:2021xfk}, this suggests that we have to dress $D_B$ with a TQFT, defining a new non-invertible operator $\mathcal{N}_B$. Formally, the partition function is still given by \eqref{egaugingnoninv}, but we need to change how we insert the defect. 

We first describe how gauge invariance is restored and how the new topological defect $\mathcal{N}_B$ is constructed; we then discuss subtleties associated with its non-invertible nature. We start by defining $\zeta^{(2)}_B\in C^2\left(X_2^B,\tfrac{\mathbb{R}}{2\pi\mathbb{Z}}\right)$ such that
\begin{align}\label{einitialdefofzetaB}
    \int_{X_2^B}\zeta^{(2)}_B=\int_{X_3}B^{(1)}\zeta^{(2)}\,.
\end{align}
The manifold $X_2^B$ is the Poincar\'e dual of $B^{(1)}$. See \eqref{eexplicitinducedcochain} for details on a specific construction. We can then define the following theory
\begin{align}\label{egaugenoninv}
    \tilde{\mathcal{Z}}_{a,B}=\underset{\phi\in C^0_A(X_2^B)}{\sum}e^{-i\int_{X_2^B}\zeta^{(2)}_B(a^{(1)},-\phi^{(0)})}\,.
\end{align}
Using \eqref{e3zetaidentity}, and translating the same property to $\zeta_B$ once integrated over $X_2^B$ via \eqref{einitialdefofzetaB}, one can prove that
\begin{align}
    \tilde{\mathcal{Z}}_{a+\delta\lambda_a,B}=e^{i\int_{X_2^B}\zeta^{(2)}_B(a^{(1)},\lambda_a^{(0)})}\tilde{\mathcal{Z}}_{a,B}\,.
\end{align}
The gauge variance of this partition function precisely compensates the anomaly in \eqref{etentativegaugeandshift}.
 
However, via explicit computations one can show that $\tilde{\mathcal{Z}}_{a,B}$ depends on the choice of the triangulation of $X_2^B$. Therefore, it is not the partition function of a well-defined TQFT. To overcome this issue, we stack \eqref{egaugenoninv} with a BF-theory coupling $a^{(1)}$ and $\delta\phi^{(0)}$ to a one-cochain $\hat{\eta}\in C^1_{\hat{A}}(X_2^B)$. The combined systems defines the following TQFT\footnote{A TQFT partition function computes a topological invariant quantity, hence it can be normalized by other topological invariants. For a nicer presentation of the fusion rules of $\mathcal{N}_B$, we choose to normalize $\mathcal{Z}_{a,B}$ by an additional $\tfrac{1}{|H^0_A|}$ factor. The normalization of a defect is fixed by how it twists the Hilbert space \cite{Shao:2023gho}, which implies that the fusion $D_aD_b=\sum_c N_{ab}^{\ \ c}D_c$ is such that $N_{ab}^{\ \ c}\in\mathbb{N}$. Here, however, we only focus on the topological nature of $\mathcal{N}_B$.}
\begin{align}\label{enoninvtheory}
    \mathcal{Z}_{a,B}=\frac{1}{|C^1_{\hat{A}}(X_2^B)||H^0_A(X_2^B)|}\underset{\phi\in C_A^0(X_2^B),\,\hat{\eta}\in C^1_{\hat{A}}(X_2^B)}{\sum}e^{-i\int_{X_2^B}\zeta^{(2)}_B(a^{(1)},-\phi^{(0)})+\hat{\eta}^{(1)}\left(a^{(1)}-\delta\phi^{(0)}\right)}\,,
\end{align}
which still satisfies
\begin{align}\label{egaugavariancetqft}
\mathcal{Z}_{a+\delta\lambda_a,B}=e^{i\int_{X_2^B}\zeta^{(2)}_B(a^{(1)},\lambda_a^{(0)})}\mathcal{Z}_a(X_2^B)\,,
\end{align}
and it is topological (see \appref{agaugevarth}). We therefore define the following non-invertible defect
\begin{align}\label{eq.nnoninvertible}
\mathcal{N}_B=\mathcal{Z}_{a,B}D_B(a^{(1)})\,,
\end{align}
which twists the partition function $Z^A$ in \eqref{egaugingnoninv} in the following way
\begin{align}\label{etentativereprnb}
Z^A[\hat{A}^{(2)},B^{(1)}]=\frac{1}{|H^0_A|}\underset{a\in H^1_A}{\sum}e^{i\int_{X_3}a^{(1)}\hat{A}^{(2)} }\mathcal{Z}_{a,B}Z[a^{(1)},B^{(1)}]\,,
\end{align}
finally producing a gauge-invariant quantity.

Additional comments are however required in order to state this result more precisely. As discussed above, for non-invertible symmetries we use a one-cocycle $B^{(1)}$ to refer to a single group element and a single two-cycle, to concisely identify the type and position of the $\mathcal{N}_B$-defect. However, we cannot use it to construct networks. The general way in which we can twist the partition function with insertions $\{\mathcal{N}_{B_i}\}$ is the following
\begin{align}\label{ereprnb}
    Z^A\left[\hat{A}^{(2)},\{B^{(1)}_i\}\right]=\frac{1}{|H^0_A|}\underset{a\in H^1_A}{\sum}e^{i\int_{X_3}\hat{A}^{(2)} a^{(1)}}\underset{i}{\prod}\mathcal{Z}_{a,B_i}Z\bigg[a^{(1)},\underset{i}{\sum}B^{(1)}_i\bigg]\,.
\end{align}
We can replace any composition of $\mathcal{N}_B$-defects by the resulting fusion rules, we present below in \eqref{enoninvrules}-\eqref{enoninvrules2}, as this is by construction equivalent to manipulating the expression above (see \appref{afusionrules} for more details).

It is sometimes possible to prove that the choice of a topological theory, needed to cancel the non-gauge invariance of a symmetry defect, is univocal modulo decoupled TQFT sectors. For instance, for certain duality defects in 4d \cite{Kaidi:2021xfk}, the corresponding unique TQFT is the $\mathcal{A}^{n,p}$ theory \cite{Hsin:2018vcg, Kaidi:2021xfk}. As we do not have such a proof in this case, we do not claim that \eqref{enoninvtheory} is the unique choice, however it reproduces the fusion rules we expect. 

In \appref{afusionrules}, we explicitly compute the fusion rules involving the non-invertible defect. Here we only present the results
\begin{align}\label{enoninvrules}
    \begin{split}
    &\mathcal{N}_B^\dag=\mathcal{N}_{-B}\,,\\[0.5em]
    &\mathcal{N}_{B_1}\mathcal{N}_{B_2}=\mathcal{N}_{B_1+B_2}\,,\ \ \ \text{if}\ B_1\neq-B_2\,,\\[0.2em]
    &\mathcal{N}_B\mathcal{N}_{-B}=\frac{\chi_A(X_2^B)}{|H^0_A(X_2^B)||H^0_{\hat{A}}(X_2^B)|}\underset{\hat{a}_B\in H^1(G_a^{(0)},X_2^B)}{\sum}D_{\hat{a}_B}\,.
    \end{split}
\end{align}
The sum in the last relation is over the insertion of all the 1-form $\hat{G}_A^{(1)}$-defects, in all independent 1-cycles inside the 2-manifold $X_2^B$ defined by $B^{(1)}$. This can be understood as \emph{condensation defect} implementing the one-gauging of the 1-form symmetry $\hat{G}_A^{(1)}$ \cite{Roumpedakis:2022aik}. Two defects commute and do not compose unless they live in the same 2-cycle. This composition, except for the last relation, follows the same rule as the original group $G_B^{(0)}$. Finally, if the 1-form symmetry defect $D_{\hat{A}}$ lives on the same manifold as $\mathcal{N}_B$, we find
\begin{align}\label{enoninvrules2}
    \mathcal{N}_B D_{\hat{A}}=D_{\hat{A}}\mathcal{N}_B=\mathcal{N}_B\,,
\end{align}
which means that $D_{\hat{A}}$ is absorbed by $\mathcal{N}_B$.

\paragraph{Gauging $\hat{G}_A^{(1)}$.} 
Gauging the quantum 1-form symmetry in $\mathcal{T}/G_A^{(0)}$ has the effect of decoupling the TQFT \eqref{enoninvtheory} that we stacked onto the system. Since our aim is to perform a topological operation that returns the original theory, we shall, by a slight abuse of terminology, refer to gauging in this case as the combined operation of gauging $\hat{G}_A^{(1)}$ and projecting out the additional topological degrees of freedom introduced by the stacked TQFT. With this convention, one readily verifies that
\begin{align}
\begin{split}\label{egaugingbackninv}
    Z^{\hat{A},A}[\hat{\hat{A}}^{(1)},\{B_i^{(1)}\}]&\doteq\frac{H^0_{\hat{A}}}{H^1_{\hat{A}}}\sum_{\hat{a}\in H^2_{\hat{A}}}e^{i\int_{X_3}\hat{\hat{A}}^{(1)}\hat{a}^{(2)}}\underset{i}{\prod}\mathcal{Z}_{-{\hat{\hat{A}}},B_i}^{-1}Z^A\left[\hat{a}^{(2)},\{B^{(1)}_i\}\right]\\
&=\chi_A^{-1} Z[-\hat{\hat{A}}^{(1)},\underset{i}{\sum}B^{(1)}_i]\,.
\end{split}
\end{align}
As expected, gives the partition function of the parent theory $\mathcal{T}$.

\section{Duality Defect}\label{sec.3}
\begin{figure}[h!]\centering
    \begin{tikzpicture}[scale=2.0, every node/.style={font=\small}]
    
    % The three theories at the vertices of an equilateral triangle
    \node[draw, rounded corners, minimum size=12mm, fill=blue!5] (TA) at (90:2.4) {$\mathcal{T}_A$};
    \node[draw, rounded corners, minimum size=12mm, fill=blue!5] (TB) at (210:2.4) {$\mathcal{T}_B$};
    \node[draw, rounded corners, minimum size=12mm, fill=blue!5] (TC) at (330:2.4) {$\mathcal{T}_C$};
    
    % Forward arrows (clockwise): solid
    % T_A -> T_C (right side)
    \draw[->, thick, >=stealth] 
        (TA) to[bend left=22] 
        node[midway, sloped, above] {\footnotesize $\mathbb{G}_{\hat{A},C}$\,\,gauging} 
        (TC);
    % T_C -> T_B (bottom)
    \draw[->, thick, >=stealth] 
        (TC) to[bend left=22] 
        node[midway, below] {\footnotesize $\mathbb{G}_{\hat{C},B}$\,\,gauging} 
        (TB);
    % T_B -> T_A (left side)
    \draw[->, thick, >=stealth] 
        (TB) to[bend left=22] 
        node[midway, sloped, above] {\footnotesize $\mathbb{G}_{\hat{B},A}$\,\,gauging} 
        (TA);
    
    % Backward arrows (counter-clockwise): dashed
    % T_C -> T_A (right side, inner)
    \draw[->, thick, >=stealth, dashed, gray!70!black] 
        (TC) to[bend left=22] 
        node[midway, sloped, below] {\footnotesize gauge $\hat{A}^{(1)}+C$} 
        (TA);
    % T_B -> T_C (bottom, inner)
    \draw[->, thick, >=stealth, dashed, gray!70!black] 
        (TB) to[bend left=22] 
        node[midway, above] {\footnotesize gauge $\hat{C}^{(1)}+B$} 
        (TC);
    % T_A -> T_B (left side, inner)
    \draw[->, thick, >=stealth, dashed, gray!70!black] 
        (TA) to[bend left=22] 
        node[midway, sloped, below] {\footnotesize gauge $\hat{B}^{(1)}+A$} 
        (TB);
    
    \end{tikzpicture}
    \caption{The three theories $\mathcal{T}_A$, $\mathcal{T}_B$, $\mathcal{T}_C$, obtained from the parent theory $\mathcal{T}$ by gauging $G_A^{(0)}$, $G_B^{(0)}$, or $G_C^{(0)}$ respectively, are connected by gauging the corresponding 2-group structure $\mathbb{G}_{\hat{X},Y}$ (solid arrows). The reverse direction (dashed arrows) is realized by first gauging the the 1-form symmetry $\hat{X}^{(1)}$ and then the 0-form symmetry $Y^{(0)}$. Equivalently, the reverse direction can be obtained by gauging twice a 2-group.}
    \label{fig.cycletheories}
\end{figure}
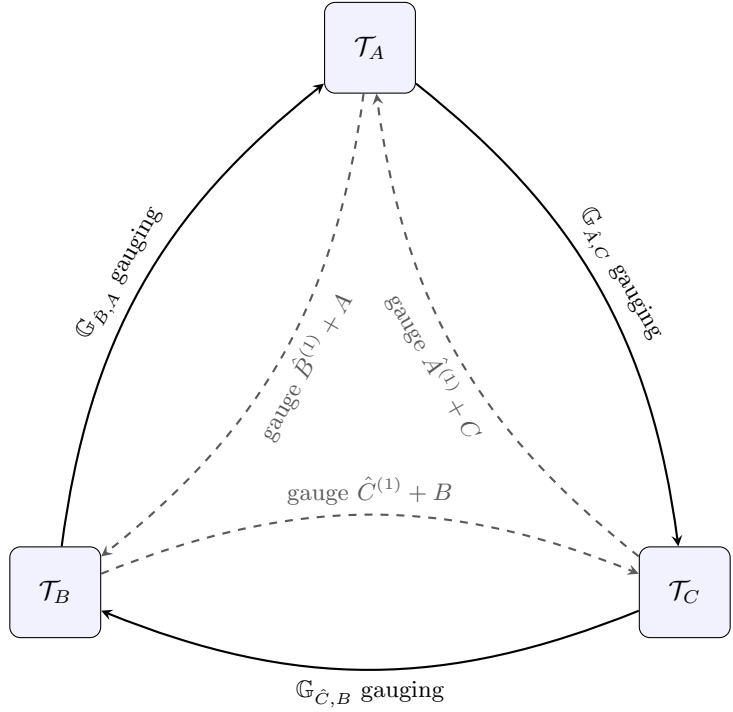

In this section we construct a non-invertible duality defect, for a 2-group symmetry, via half-spacetime gauging. This a generalization of the results obtained for three-dimensional theories in \cite{Choi:2024rjm,Cui:2024cav}, where anomaly free theories have been considered. To this end, we consider the parent theory $\mathcal{T}$, now with $G_A^{(0)}\times G_B^{(0)}\times G_C^{(0)}$ symmetry and the following anomaly theory
\begin{equation}\label{eq.4danomT}
    S_\text{anomaly}=\int_{Y_4}A^{(1)} C^{(1)*}\beta_{AC}+B^{(1)} A^{(1)*}\beta_{BA}+C^{(1)} B^{(1)*}\beta_{CB}\,.
\end{equation}
We denote by $\mathcal{T}_A$ the theory obtained from $\mathcal{T}$ upon gauging $G_A^{(0)}$. Computations analogous to those in \autoref{sec.2.1} and \autoref{sec.2.2} show that $\mathcal{T}_A$ possesses a 2-group structure, between $G_C^{(0)}$ and $\hat{G}^{(1)}_A$, and a non-invertible symmetry descending from the original $G_B^{(0)}$ (whose fusion rules are identical to \eqref{enoninvrules} and include the contributions from $\hat{G}_A^{(1)}$). Gauging, respectively, $G_B^{(0)}$ and $G_C^{(0)}$, defines the theories $\mathcal{T}_B$ and $\mathcal{T}_C$. These theories are connected by gauging the entire 2-group, see \autoref{fig.cycletheories}.
Whenever $G^{(0)}=G_A^{(0)}=G_B^{(0)}=G_C^{(0)}$ and $\beta=\beta_{AC}=\beta_{BA}=\beta_{CB}$, the theories $\mathcal{T}_A$, $\mathcal{T}_B$, $\mathcal{T}_C$ may, in principle, be mutually dual. If this is shown to be the case, that is
\begin{equation}\label{eq.permutation}
    \cT_A\simeq\cT_A/\mathbb{G}_{\hat{A},C}=\cT_C,\qquad \cT_C\simeq\cT_C/\mathbb{G}_{\hat{C},B}=\cT_B\,,\qquad
    \cT_B\simeq\cT_B/\mathbb{G}_{\hat{B},A}=\cT_A\,,
\end{equation}
the duality defect $\mathcal{D}$ implementing the 2-group half-gauging gives rise to a non-invertible symmetry. This implies that the theory $\mathcal{T}_A$ enjoys an $S_3$ symmetry, where $S_3$ is the group of permutations acting on the three symmetry groups of $\mathcal{T}_A$.
 
We take $\mathcal{T}_A$ as the reference theory for computing the fusion rules. In this theory, $D_{\hat{A}}$ is the 1-form symmetry forming a 2-group with $D_C$, while $\mathcal{N}_B$ is the non-invertible defect. In \autoref{sdualitycondenstation} we prove the following non-trivial fusion rules involving the duality defect
\begin{align}\label{efusiondualitydefect}
    \begin{split}
    &D_{\hat{A}}\mathcal{D}=\mathcal{D} D_{\hat{A}}=\mathcal{D}\,,\\
    &\mathcal{N}_B\mathcal{D}=\mathcal{D}\mathcal{N}_B=\mathcal{D}\,,\\
    &\mathcal{D}\times\mathcal{D}=\chi^{-2}_{G,\geq 0}\sum_b\mathcal{N}_b\,,
    \end{split}
\end{align}
where the summation over $b$ runs over all possible types of non-invertible defects $\mathcal{N}_B$ and all 2-cycles in the manifold that supports $\mathcal{D}$. This can be interpreted as a condensation defect implementing the 1-gauging of the non-invertible symmetry \cite{Seifnashri:2024dsd}. Moreover, $\chi_{G,\geq 0}$ is the Euler counterterm associated with the region of spacetime where the half-gauging is applied. 

It also follows from the definition, and the choice of normalization, that \cite{Kaidi:2022cpf}
\begin{align}
\mathcal{D}^\dag=\chi^{2}_{G,\geq 0}\mathcal{D}\,.
\end{align}
In the following subsections we describe more concretely the relations among the partition functions, the half-gauging procedure and the fusion rules of $\mathcal{D}$.

\subsection{Transformations among $Z^A,\, Z^B,\, Z^C$}
We define the partition function of $\mathcal{T}_A$ as
\begin{align}\label{eza}
    Z^A[C^{(1)},\hat{A}^{(2)},B^{(1)}]=\frac{1}{|H_A^0|}\underset{a\in H^1_A}{\sum}e^{i\int\hat{A}^{(2)}a^{(1)}}Z[a^{(1)},B^{(1)},C^{(1)}]\,,
\end{align}
where, to lighten the notation, we write $B^{(1)}$ instead of $\{B^{(1)}\}$ for the non-invertible defect (see \eqref{ereprnb}). It is possible to show that
\begin{align}\label{eanomalytripleth}
    Z^A[C^{(1)}+\delta\lambda_C^{(0)},\hat{A}^{(2)}+\delta\lambda_{\hat{A}}^{(1)}+\zeta^{(2)}(C^{(1)},\lambda_C^{(0)}),B^{(1)}]=e^{i\int\lambda_C^{(0)} B^{(1)*}\beta}Z^A[C^{(1)},\hat{A}^{(2)},B^{(1)}]\,.
\end{align}
This phase represents a mixed 't Hooft anomaly between the 2-group $\mathbb{G}_{\hat{A},C}$ and the non-invertible symmetry generated by $\mathcal{N}_B$. This anomaly can be anticipated from the last term in the anomaly theory of the parent theory, \eqref{eq.4danomT}. More precisely, for an individual $\mathcal{N}_B$ insertion labelled by the cocycle $B^{(1)}$, the corresponding four-dimensional anomaly theory is
\begin{equation}
    S_\text{anomaly}=\int_{Y_4} C^{(1)} B^{(1)*}\beta_{CB}.
\end{equation}
Its boundary variation reproduces the anomalous transformation in \eqref{eanomalytripleth}. We stress that $B^{(1)}$ is used here only to specify a single non-invertible defect insertion; a general network of $\mathcal{N}_B$ defects cannot be described by an ordinary group-valued background field and should instead be formulated directly in terms of the
corresponding defect network. 

Linking the theories $\mathcal{T}_A,\mathcal{T}_B,\mathcal{T}_C$ requires gauging the 2-group structure, as shown in \autoref{fig.cycletheories}. In addition, one must stack and subsequently project out the auxiliary TQFTs used to define the non-invertible symmetries (as in \eqref{ereprnb}-\eqref{egaugingbackninv}). Furthermore, we require charge conjugation due to the minus sign that appears in the source for $A^{(1)}$, see \eqref{egaugingbackninv}. Concretely, defining $Z^B$ and $Z^C$ analogously to $Z^A$ in \eqref{eza}, one can indeed prove that
\begin{align}
    \chi_A^{-1} Z^C[B^{(1)},\hat{C}^{(2)},A^{(1)}]=\frac{|H^0_{\hat{A}}|}{|H_{\hat{A}}^1||H^0_C|}\hspace{-0.5 cm}\underset{\underset{\hat{a}\in H^2_{\hat{A},\delta\hat{a}=c^*\beta_{AC}}}{c\in H^1_C}}{\sum}\hspace{-0.7 cm}e^{i\int \hat{C}^{(2)}c^{(1)}-A^{(1)}\hat{a}^{(2)}}\frac{\mathcal{Z}_{c,A}}{\mathcal{Z}_{A,B}}Z^A[c^{(1)},\hat{a}^{(2)},B^{(1)}]\,,
\end{align}
with similar relations for the other partition functions. The mixed anomaly in \eqref{eanomalytripleth} does not obstruct the gauging of the 2-group, since its anomalous phase vanishes in the absence of $\mathcal{N}_B$ insertions. Under the gauging of $\mathbb{G}_{\hat{A},C}$, the non-invertible symmetry generated by $\mathcal{N}_B$ is mapped to the invertible zero-form symmetry $G_B^{(0)}$, which combines with the quantum one-form symmetry $\hat{G}_C^{(1)}$ into the 2-group of the resulting theory $\mathcal{T}_C$. Moreover, $\mathcal{T}_C$ possesses the cyclically related mixed anomaly between this 2-group and the non-invertible symmetry generated by $\mathcal{N}_A$, as required by anomaly matching under the proposed duality with $\mathcal{T}_A$.

Referring to \autoref{fig.cycletheories}, we can also relate $Z^A$ to $Z^B$ by gauging the 1-form symmetry, stacking a suitable TQFT, projecting out the TQFT associated with the original non-invertible symmetry and finally gauging $G^{(0)}_B$.\footnote{This procedure admits the interpretation of gauging the non-invertible symmetry $\mathcal{N}_B$ together with the 1-form symmetry. Indeed, the former cannot be gauged without also gauging the latter, which appears in the fusion rule of $\mathcal{N}_B\mathcal{N}_{-B}$ \eqref{enoninvrules}. Alternatively, this can be seen as gauging twice the 2-group symmetry, traversing the longer path around the circle of transformations in \autoref{fig.cycletheories}.} Indeed, one can show that
\begin{align}\label{egauging1formnoninv}
    \chi_A^{-1}Z^B[C^{(1)},\hat{B}^{(2)},A^{(1)}]=\frac{|H^0_{\hat{A}}|}{|H_{\hat{A}}^1||H^0_C|}\hspace{-0.5 cm}\underset{\underset{\hat{a}\in H^2_{\hat{A},\delta\hat{a}=C^*\beta_{AC}}}{b\in H^1_B}}{\sum}\hspace{-0.7 cm}e^{i\int \hat{B}^{(2)}b^{(1)}-A^{(1)}\hat{a}^{(2)}}\frac{\mathcal{Z}_{b,C}}{\mathcal{Z}_{A,b}}Z^A[c^{(1)},\hat{a}^{(2)},B^{(1)}]\,,
\end{align}
together with other two similar relations.
 
\subsection{Half-Spacetime Gauging}
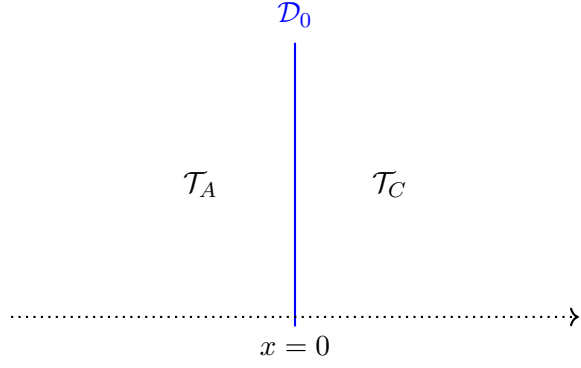
\begin{figure}[!tbp]
\centering
{\begin{tikzpicture}[baseline=80,scale=1.25]
\draw[blue, thick] (2,-0)--(2,3.0);
\draw[ thick,dotted,->] (-1,0.1) -- (5,0.1);
\node[blue] at (2,3.3) {$\mathcal{D}_0$};
\node[below] at (2,0) {$x=0$};
\node at (1,1.5) {$\mathcal{T}_A$};
\node at (3,1.5) {$\mathcal{T}_C $};
\end{tikzpicture}}
\caption{Depiction of the duality wall that implements the 2-group gauging on one side. We label the duality defect by $\mathcal{D}_0$ to denote that it is inserted at $x=0$.}
\label{fig.wei}
\end{figure}

We consider a region of the compact space $X$ to be described, locally, by $I\times Y$, with $I\subset\mathbb{R}$. We choose a point $0\in I$ and divide the manifold accordingly into two halves. We gauge a symmetry on $X_{\geq 0}$, which has boundary $\partial X_{\geq 0}\cong Y$. Half-gauging requires summing over connections on the half-spacetime with Dirichlet boundary conditions; we refer the reader to \appref{ahalfspacetimegauging} for details. See \autoref{fig.wei} for an illustration of the setup.

Similarly to \eqref{ehalfgauging}, we can express the partition function in presence of $\mathcal{D}_0$ as 
\begin{align}
    \begin{split}\label{eformuladoublegauging}
    &Z_{\mathcal{D}_0}^A[\{C^{(1)}_{\leq0},B^{(1)}_{\geq0}\},\{\hat{A}^{(2)}_{\leq0},\hat{C}^{(2)}_{\geq0}\},\{B^{(1)}_{\leq0},A^{(1)}_{\geq0}\}]\\
    &=\frac{|H^0_{\hat{A},>0}|}{|H^1_{\hat{A},>0}||H^0_{C,>0}|}\hspace{-0.5 cm}\underset{\underset{\hat{a}\in H^2_{\hat{A},>0,\delta\hat{a}=c^*\beta_{AC}}}{c\in H^1_{C,>0}}}{\sum}\hspace{-0.7 cm}e^{i\int_{X_{\geq 0}} \hat{C}^{(2)}_{\geq 0}c^{(1)}_{>0}-A^{(1)}_{\geq 0}\hat{a}^{(2)}_{>0}}\frac{\mathcal{Z}_{c_{>0},A_{\geq 0}}}{\mathcal{Z}_{A_{\geq 0},B_{\geq 0}}}\\
    &\hspace{4.7 cm}Z^A[C^{(1)}_{\leq 0}+c^{(1)}_{>0},\hat{A}^{(2)}_{\leq0}+\hat{a}^{(2)}_{>0},B^{(1)}_{\leq 0}+B^{(1)}_{\geq 0}]
    \end{split}
\end{align}
and, similarly as in \eqref{edualityabs}, one can prove that
\begin{align}
    \begin{split}
    &D_{\hat{A}}\mathcal{D}_0=\mathcal{D}_0 D_{\hat{C}}=\mathcal{D}_0\,,
    \end{split}
\end{align}
which justifies the first of the fusion rules in \eqref{efusiondualitydefect}. A straightforward calculation, assuming that \eqref{eq.permutation} holds, would also prove the second equation in \eqref{efusiondualitydefect}. Using the properties listed in \eqref{efundamentalfourierrelative}, we can write \eqref{eformuladoublegauging} as
\begin{align}
    \begin{split}\label{eformuladoublegauging2}
    &Z_{\mathcal{D}_0}^A[\{C^{(1)}_{\leq0},B^{(1)}_{\geq0}\},\{\hat{A}^{(2)}_{\leq0},\hat{C}^{(2)}_{\geq0}\},\{B^{(1)}_{\leq0},A^{(1)}_{\geq0}\}]\\
    &=\frac{1}{|H^0_{C,>0}||C^1_{A,\geq 0}|\chi_{A,\geq0}}\underset{\underset{\hdots}{c\in H^1_{C,>0}}}{\sum}\hspace{-0.2 cm}e^{i\int_{X_{\geq 0}} \hat{C}^{(2)}_{\geq 0}c^{(1)}_{>0}-A^{(1)}_{\geq 0}\hat{a}^{(2)}_{>0}+\phi^{(0)}_{A,\geq 0}(\delta \hat{a}^{(2)}-c^{(1)*}_{>0}\beta_{AC})}\frac{\mathcal{Z}_{c_{>0},A_{\geq 0}}}{\mathcal{Z}_{A_{\geq 0},B_{\geq 0}}}\\
    &\hspace{5 cm}Z^A[C^{(1)}_{\leq 0}+c^{(1)}_{>0},\hat{A}^{(2)}_{\leq0}+\hat{a}^{(2)}_{>0},B^{(1)}_{\leq 0}+B^{(1)}_{\geq 0}]\,.
    \end{split}
\end{align}
The summation runs over all lower-case fields and, unless otherwise specified, over cochains. The subscript specifies the region on which the cochain is defined, e.g. $> 0$ and $\geq 0$ indicates, respectively, whether the cochain vanishes on the boundary or not.

\subsection{Duality Defect Fusion Rules}\label{sdualitycondenstation}
Assuming that the theories on the two sides of the wall are dual, the defect $\mathcal{D}_0$ becomes a symmetry and we can rewrite \eqref{eformuladoublegauging2} as
\begin{align}
    \begin{split}\label{eformuladoublegaugingdual}
    &Z_{\mathcal{D}_0}^A[\{C^{(1)}_{\leq0},C^{(1)}_{\geq0}\},\{\hat{A}^{(2)}_{\leq0},\hat{A}^{(2)}_{\geq0}\},\{B^{(1)}_{\leq0},B^{(1)}_{\geq0}\}]\\
    &=\frac{1}{|H^0_{G,>0}||C^1_{G,\geq 0}|\chi_{G,\geq0}}\underset{\underset{\hdots}{c\in H^1_{G,>0}}}{\sum}\hspace{-0.2 cm}e^{i\int_{X_{\geq 0}} \hat{A}^{(2)}_{\geq 0}c^{(1)}_{>0}-B^{(1)}_{\geq 0}\hat{a}^{(2)}_{>0}+\phi^{(0)}_{A,\geq 0}(\delta \hat{a}^{(2)}-c^{(1)*}_{>0}\beta)}\frac{\mathcal{Z}_{c_{>0},B_{\geq 0}}}{\mathcal{Z}_{B_{\geq 0},B_{\geq 0}}}\\
    &\hspace{5 cm}Z^A[C^{(1)}_{\leq 0}+c^{(1)}_{>0},\hat{A}^{(2)}_{\leq0}+\hat{a}^{(2)}_{>0},B^{(1)}_{\leq 0}+C^{(1)}_{\geq 0}]\,,
    \end{split}
\end{align}
where we recall that, in the duality case, $G^{(0)}=G_A^{(0)}=G_B^{(0)}=G_C^{(0)}$ and $\beta=\beta_{AC}=\beta_{BA}=\beta_{CB}$. 

\begin{figure}[!tbp]
\centering
\begin{tikzpicture}[baseline=50,scale=1.25]
\draw[blue, thick] (2,-0)--(2,3);
\draw[blue, thick] (4,-0)--(4,3);
\node[blue] at (2,3.3) {$\cD_0$};
\node[blue] at (4,3.3) {$\cD_{\epsilon}$};
\node[below] at (2,0) {$x=0$};
\node[below] at (4,0) {$x=\epsilon$};
\draw[thick,dotted,->] (-1,0.1) -- (7,0.1);
\node at (1,1.5) {$\mathcal{T}_A$};
\node at (3,1.5) {$\mathcal{T}_A$};
\node at (5,1.5) {$\mathcal{T}_A$};
\end{tikzpicture}
\caption{Fusion of two duality defects $\mathcal{D}_0$ and $\mathcal{D}_\epsilon$.}
\label{fig.wei2}
\end{figure}
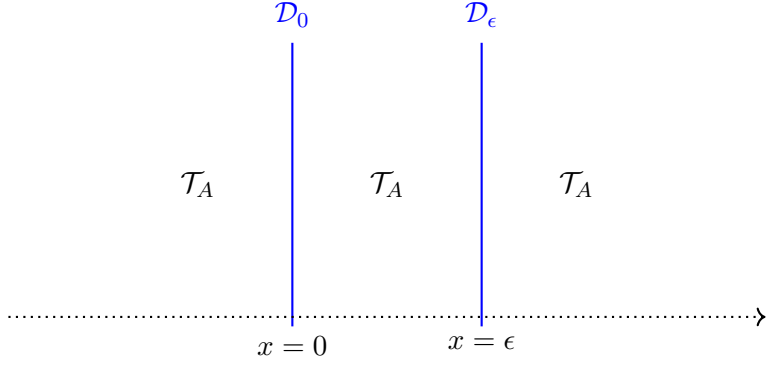

We then introduce a second duality defect, displaced from the first by $\varepsilon$, with which it can fuse. This setup is shown in \autoref{fig.wei2}.
The partition function then reads
\begin{align}
    \begin{split}
    Z_{\mathcal{D}_0\mathcal{D}_\varepsilon}^A=\kappa_1 \sum_{\hdots} &e^{i\int_{X_{\geq\varepsilon}}\hat{A}^{(2)}_{\geq\varepsilon}c^{(1)}_{>\varepsilon}-B^{(1)}_{\geq\varepsilon}\hat{a}^{(2)}_{>\varepsilon}+\phi_{A,\geq\varepsilon}(\delta\hat{a}^{(2)}_{>\varepsilon}-c_{>\varepsilon}^{(1)*}\beta)+\phi_{\hat{C},\geq \varepsilon}^{(1)}\delta c^{(1)}_{>\varepsilon}}\\
    &e^{i\int_{X_{\geq0}}(\hat{A}^{(2)}_{[0,\varepsilon]}+\hat{a}^{(2)}_{>\varepsilon})\tilde{c}^{(1)}_{>0}-(B^{(1)}_{[0,\varepsilon]}+C_{\geq \varepsilon})\tilde{\hat{a}}^{(2)}_{>0}+\tilde{\phi}_{A,\geq0}(\delta\tilde{\hat{a}}^{(2)}_{>0}-\tilde{c}_{>0}^{(1)*}\beta)+\tilde{\phi}_{\hat{C},\geq 0}^{(1)}\delta \tilde{c}^{(1)}_{>0}}\\
    &\frac{\mathcal{Z}_{\tilde{c}_{(0,\varepsilon)},B_{[0,\varepsilon]}}\mathcal{Z}_{\tilde{c}_{\geq \varepsilon},{C_{\geq \varepsilon}}}\mathcal{Z}_{c_{>\varepsilon},B_{\geq\varepsilon}}}{\mathcal{Z}_{B_{[0,\varepsilon]},C_{[0,\varepsilon]}}\mathcal{Z}_{C_{\geq\varepsilon},c_{>\varepsilon}}\mathcal{Z}_{B_{\geq\varepsilon},C_{\geq\varepsilon}}}Z^A[C_{\leq 0}+\tilde{c}^{(1)}_{>0},\hat{A}^{(2)}_{\leq 0}+\tilde{\hat{a}}^{(2)}_{>0},B_{\leq0}^{(1)}+C^{(1)}_{[0,\varepsilon]}+c^{(1)}_{>\varepsilon}]\,,
    \end{split}
\end{align}
where Lagrange multipliers have been introduced, so as to work entirely with cochains, and
\begin{align}
    \kappa_1=\frac{1}{|H^0_{G,>0}||C^1_{G,\geq 0}||C^1_{\hat{G},\geq 0}|\chi_{G,\geq0}|H^0_{G,>\varepsilon}||C^1_{G,\geq \varepsilon}||C^1_{\hat{G},\geq \varepsilon}|\chi_{G,\geq\varepsilon}}\,.
\end{align}
When necessary we split cochains in different subspaces, e.g. $\phi_{\geq 0}=\phi_{[0,\varepsilon]}+\phi_{>\varepsilon}$, where we define the new cochains as being equal to the identity outside their defining region (e.g. $\phi_{[0,\varepsilon]}=\mathds{1}$ outside $X_{[0,\varepsilon]}$). We have also used the fact that the TQFT partition function factorises \eqref{enoninvtheory}, e.g.
\begin{align}
    \mathcal{Z}_{\tilde{c}_{>0},B_{[0,\varepsilon]}+C_{\geq \varepsilon}}=\mathcal{Z}_{\tilde{c}_{(0,\varepsilon)},B_{[0,\varepsilon]}}\mathcal{Z}_{\tilde{c}_{\geq \varepsilon},{C_{\geq \varepsilon}}}\,.
\end{align}
Combining these observations, we can rewrite $Z_{\mathcal{D}_0\mathcal{D}_\varepsilon}^A$ as
\begin{align}
    \begin{split}
    Z_{\mathcal{D}_0\mathcal{D}_\varepsilon}^A=\kappa_1 \sum_{\hdots} &e^{i\int_{X_{\geq\varepsilon}}\hat{a}^{(2)}_{>\varepsilon}(\tilde{c}^{(1)}_{\geq\varepsilon}-B^{(1)}_{\geq\varepsilon}-\delta\phi^{(0)}_{A,\geq\varepsilon})+\hat{A}^{(2)}_{\geq\varepsilon}c^{(1)}_{>\varepsilon}-\phi_{A,\geq\varepsilon}c_{>\varepsilon}^{(1)*}\beta+\phi_{\hat{C},\geq \varepsilon}^{(1)}\delta c^{(1)}_{>\varepsilon}}\\
    &e^{i\int_{X_{\geq0}}\hat{A}^{(2)}_{[0,\varepsilon]}\tilde{c}^{(1)}_{>0}-(B^{(1)}_{[0,\varepsilon]}+C_{\geq \varepsilon})\tilde{\hat{a}}^{(2)}_{>0}+\tilde{\phi}_{A,\geq0}(\delta\tilde{\hat{a}}^{(2)}_{>0}-\tilde{c}_{>0}^{(1)*}\beta)+\tilde{\phi}_{\hat{C},\geq 0}^{(1)}\delta \tilde{c}^{(1)}_{>0}}\\
    &\frac{\mathcal{Z}_{\tilde{c}_{(0,\varepsilon)},B_{[0,\varepsilon]}}\mathcal{Z}_{\tilde{c}_{\geq \varepsilon},{C_{\geq \varepsilon}}}\mathcal{Z}_{c_{>\varepsilon},B_{\geq\varepsilon}}}{\mathcal{Z}_{B_{[0,\varepsilon]},C_{[0,\varepsilon]}}\mathcal{Z}_{C_{\geq\varepsilon},c_{>\varepsilon}}\mathcal{Z}_{B_{\geq\varepsilon},C_{\geq\varepsilon}}}Z^A[C_{\leq 0}+\tilde{c}^{(1)}_{>0},\hat{A}^{(2)}_{\leq 0}+\tilde{\hat{a}}^{(2)}_{>0},B_{\leq0}^{(1)}+C^{(1)}_{[0,\varepsilon]}+c^{(1)}_{>\varepsilon}]\,.
    \end{split}
\end{align}
We then shift the summation variables $\tilde{c}_{>0}^{(1)}$ and $\tilde{\hat{a}}^{(2)}_{>0}$ according to $\tilde{c}_{>0}^{(1)}\rightarrow \tilde{c}_{>0}^{(0)}+\delta\phi_{A,\geq\varepsilon}^{(0)},\,\tilde{\hat{a}}^{(2)}_{>0}\rightarrow\tilde{\hat{a}}^{(2)}_{>0}+\zeta^{(2)}(\tilde{c}^{(1)}_{\geq\varepsilon},\phi^{(0)}_{A,\geq\varepsilon})$. Using \eqref{egaugavariancetqft}-\eqref{eanomalytripleth}, we can see that the sum becomes independent of $\phi_{A,\geq\varepsilon}$ and $\tilde{\phi}^{(1)}_{\hat{C},>\varepsilon}$. We may then remove the corresponding summations, picking up a factor of $|C^0_{G,\geq\varepsilon}||C^1_{\hat{G},>\varepsilon}|$. 

Summing over $\hat{a}^{(2)}_{>\varepsilon}$ produces a factor $|C^2_{\hat{G},>\varepsilon}|$ and enforces the constraint $\tilde{c}^{(1)}_{\geq \varepsilon}=B^{(1)}_{\geq\varepsilon}$. This brings the partition function into the form
\begin{align}
    \begin{split}
    Z_{\mathcal{D}_0\mathcal{D}_\varepsilon}^A=\kappa_2 \sum_{\hdots} &e^{i\int_{X_{[0,\varepsilon]}}\hat{A}^{(2)}_{[0,\varepsilon]}\tilde{c}^{(1)}_{(0,\varepsilon)}-B^{(1)}_{[0,\varepsilon]}\tilde{\hat{a}}^{(2)}_{(0,\varepsilon)}+\tilde{\phi}_{A,[0,\varepsilon]}(\delta\tilde{\hat{a}}^{(2)}_{(0,\varepsilon)}-\tilde{c}_{(0,\varepsilon)}^{(1)*}\beta)+\tilde{\phi}_{\hat{C},[0,\varepsilon]}^{(1)}\delta \tilde{c}^{(1)}_{(0,\varepsilon)}}\frac{\mathcal{Z}_{\tilde{c}_{(0,\varepsilon)},B_{[0,\varepsilon]}}}{\mathcal{Z}_{B_{[0,\varepsilon]},C_{[0,\varepsilon]}}}\\
    &e^{i\int_{X_{\geq\varepsilon}}\hat{A}^{(2)}_{\geq\varepsilon}c^{(1)}_{>\varepsilon}-C^{(1)}_{\geq \varepsilon}\tilde{\hat{a}}^{(2)}_{\geq\varepsilon}+\phi_{\hat{C},\geq \varepsilon}^{(1)}\delta c^{(1)}_{>\varepsilon}+\tilde{\phi}_{A,>\varepsilon}(\delta\tilde{\hat{a}}^{(2)}_{\geq\varepsilon}-B^{(1)*}_{\geq\varepsilon}\beta)}\frac{\mathcal{Z}_{c_{>\varepsilon},B_{\geq\varepsilon}}}{\mathcal{Z}_{C_{\geq\varepsilon},c_{>\varepsilon}}}\\
    &Z^A[C_{\leq 0}+\tilde{c}^{(1)}_{>0},\hat{A}^{(2)}_{\leq 0}+\tilde{\hat{a}}^{(2)}_{>0},B_{\leq0}^{(1)}+C^{(1)}_{[0,\varepsilon]}+c^{(1)}_{>\varepsilon}]\,,
    \end{split}
\end{align}
with
\begin{align}
    \kappa_2=\frac{|C^0_G(\partial X_{\geq\varepsilon})|}{|C^0_{G,>0}||C^1_{G,\geq 0}||C^1_{\hat{G},\geq0}|\chi_{G,\geq0}\chi_{G,\geq\varepsilon}|C^1_{\hat{G}}(\partial X_{\geq \varepsilon})|}\,.
\end{align}
Up to a relabeling of the sources in \eqref{eformuladoublegaugingdual}, we recognize in the region $X_{\geq \varepsilon}$ the gauging of the 1-form and non-invertible symmetries described in \eqref{egauging1formnoninv}.\footnote{We have a summation over $\tilde{\hat{a}}^{(2)}_{\geq\varepsilon}$, thus in the absolute cohomology. This because of the constraint $\delta\tilde{\hat{a}}^{(2)}_{\geq\varepsilon}=B^{(1)*}_{\geq \varepsilon}\beta$ coming from the external source $B^{(1)}_{\geq\varepsilon}$, which belongs to the absolute cohomology.} Under the assumption of duality among $\mathcal{T}_A$, $\mathcal{T}_B$, $\mathcal{T}_C$, the bulk in $X_{\geq\varepsilon}$ remains unchanged. More concretely, one can rewrite the expression as 
\begin{align}
    \begin{split}
    Z_{\mathcal{D}_0\mathcal{D}_\varepsilon}^A=\chi_{G,\geq\varepsilon}^{-2}\frac{|H^0_{\hat{G},(0,\varepsilon)}|}{|H^0_{G,(0,\varepsilon)}||H^1_{\hat{G},(0,\varepsilon)}|} &\hspace{-0.7 cm}\underset{\tilde{\hat{a}}^{(2)}\in H^2_{\hat{G},\delta\tilde{\hat{a}}=\tilde{c}^*\beta,(0,\varepsilon)}}{\sum_{\tilde{c}\in H^1_{G,(0,\varepsilon)}}}\hspace{-1 cm}e^{i\int_{X_{[0,\varepsilon]}}\hat{A}^{(2)}_{[0,\varepsilon]}\tilde{c}^{(1)}_{(0,\varepsilon)}-B^{(1)}_{[0,\varepsilon]}\tilde{\hat{a}}^{(2)}_{(0,\varepsilon)}}\frac{\mathcal{Z}_{\tilde{c}_{(0,\varepsilon)},B_{[0,\varepsilon]}}}{\mathcal{Z}_{B_{[0,\varepsilon]},C_{[0,\varepsilon]}}}\\
    \chi_{G,\geq\varepsilon}\frac{|H^0_{\hat{G},\geq\varepsilon}|}{|H^0_{G,\geq\varepsilon}||H^1_{\hat{G},\geq\varepsilon}|}&\hspace{-0.7 cm}\underset{\tilde{\hat{a}}^{(2)}\in H^2_{\hat{G},\delta\tilde{\hat{a}}=\tilde{c}^*\beta,\geq\varepsilon}}{\sum_{c\in H^1_{G,>\varepsilon}}} \hspace{-1 cm}e^{i\int_{X_{\geq\varepsilon}}\hat{A}^{(2)}_{\geq\varepsilon}c^{(1)}_{>\varepsilon}-C^{(1)}_{\geq \varepsilon}\tilde{\hat{a}}^{(2)}_{\geq\varepsilon}}\frac{\mathcal{Z}_{c_{>\varepsilon},B_{\geq\varepsilon}}}{\mathcal{Z}_{C_{\geq\varepsilon},c_{>\varepsilon}}}\\
    &\hspace{1 cm}Z^A[C_{\leq 0}+\tilde{c}^{(1)}_{>0},\hat{A}^{(2)}_{\leq 0}+\tilde{\hat{a}}^{(2)}_{>0},B_{\leq0}^{(1)}+C^{(1)}_{[0,\varepsilon]}+c^{(1)}_{>\varepsilon}]\,,
    \end{split}
\end{align}
where the first line collects all contributions associated with the fusion of the two duality defects in $X_{[0,\varepsilon]}$ \cite{Kaidi:2022cpf}. We obtain a sum over all ways of inserting the 1-form symmetry and the non-invertible 0-form symmetry operators,\footnote{One needs to pay close attention to how the names of the sources are relabeled from \autoref{fig.cycletheories}. In particular, to go from $\mathcal{T}_A$ to $\mathcal{T}_B$, we have $A\rightarrow C$, $C\rightarrow B$ and $B\rightarrow A$. We can then see that the exponentials, much like the one in \eqref{edefectrepresentation}-\eqref{edefectrepresentation2}, represent the operators we state here.} with intersection rules identical to those of the original 2-group structure. This, however, breaks down, as $\tilde{c}^{(1)*}\beta=0$ (since no 3-closed cocycle survive in the limit $\varepsilon \rightarrow 0$). Furthermore, by virtue of \eqref{enoninvrules2}, the 1-form symmetry is absorbed. Finally, all cohomological factors associated with the strip $(0,\varepsilon)$, in front of the sum, can be proven to reduce to unity. We therefore remain with 
\begin{align}
    \begin{split}
    Z_{\mathcal{D}_0\mathcal{D}_\varepsilon}^A=\chi_{G,\geq\varepsilon}^{-2} &\hspace{-0.2 cm}\underset{\tilde{\hat{a}}^{(2)}\in H^2_{\hat{G},(0,\varepsilon)}}{\sum_{\tilde{c}\in H^1_{G,(0,\varepsilon)}}} \hspace{-0.4 cm}e^{-i\int_{X_{[0,\varepsilon]}}B^{(1)}_{[0,\varepsilon]}\tilde{\hat{a}}^{(2)}_{(0,\varepsilon)}}\frac{\mathcal{Z}_{\tilde{c}_{(0,\varepsilon)},B_{[0,\varepsilon]}}}{\mathcal{Z}_{B_{[0,\varepsilon]},C_{[0,\varepsilon]}}}\\
    &\chi_{G,\geq\varepsilon}\frac{|H^0_{\hat{G},\geq\varepsilon}|}{|H^0_{G,\geq\varepsilon}||H^1_{\hat{G},\geq\varepsilon}|}\hspace{-0.6 cm}\underset{\tilde{\hat{a}}^{(2)}\in H^2_{\hat{G},\delta\tilde{\hat{a}}=\tilde{c}^*\beta,\geq\varepsilon}}{\sum_{c\in H^1_{G,>\varepsilon}}} \hspace{-0.8 cm}e^{i\int_{X_{\geq\varepsilon}}\hat{A}^{(2)}_{\geq\varepsilon}c^{(1)}_{>\varepsilon}-C^{(1)}_{\geq \varepsilon}\tilde{\hat{a}}^{(2)}_{\geq\varepsilon}}\frac{\mathcal{Z}_{c_{>\varepsilon},B_{\geq\varepsilon}}}{\mathcal{Z}_{C_{\geq\varepsilon},c_{>\varepsilon}}}\\
    &\hspace{5 cm}Z^A[C_{\leq 0}+\tilde{c}^{(1)}_{>0},\hat{A}^{(2)}_{\leq 0}+\tilde{\hat{a}}^{(2)}_{>0},B_{\leq0}^{(1)}+C^{(1)}_{[0,\varepsilon]}+c^{(1)}_{>\varepsilon}]\,.
    \end{split}
\end{align}
We can notice in the first row that, apart form a denominator that removes spurious TQFT degrees of freedom, we obtain precisely the last fusion rule in \eqref{efusiondualitydefect}.

\section{Examples}\label{sec.4}
We now present examples of theories admitting a non-invertible duality defect of the type constructed in \autoref{sec.3}, arising from the half-spacetime gauging of a 2-group. All the theories considered are product theories, generalizing those of \cite{Choi:2024rjm,Cui:2024cav}, constructed by taking the product of a seed theory $\mathcal{Q}$. In these cases, half-spacetime gauging of the 2-group acts as a permutation of the factors of the product theory. We also allow for interactions between the factors. Note that, to obtain the anomaly structure \eqref{eq.4danomT} required for a 2-group to arise upon gauging the 0-form symmetry, we stack the product theory with a TQFT providing the desired anomaly.

\subsection{$U(1)\times U(1)\times U(1)$ gauge theory}\label{sec.4.1}
Consider a 3d gauge theory $\mathcal{T}$ with gauge group $U(1)_A\times U(1)_B\times U(1)_C$. Its symmetry content consists of three copies of the electric and magnetic symmetries $U(1)^{(1)}_e\times U(1)^{(0)}_m$, one for each $U(1)$ gauge factor. We introduce an anomaly theory for the magnetic 0-form symmetries $U(1)_{m,A}\times U(1)_{m,B}\times U(1)_{m,C}$ of the form \eqref{eq.4danomT}.

Gauging a $\mathbb{Z}_{N,m,A}^{(0)}$ subgroup of the magnetic 0-form symmetry of the first factor produces a theory $\mathcal{T}_A$ with the following structure:
\begin{itemize}
    \item The electric 1-form symmetry of the first factor $\mathbb{Z}_{N,e,A}^{(1)}$, together with the magnetic 0-form symmetry of the third factor $\mathbb{Z}_{N,m,C}^{(0)}$, form a 2-group.
    \item The magnetic 0-form symmetry of the second factor $\mathbb{Z}_{N,m,B}^{(0)}$ becomes non-invertible.
\end{itemize}
Recall that gauging a $\mathbb{Z}_{N,m}^{(0)}$ magnetic symmetry for a $U(1)$ gauge factor rescales the gauge coupling $e_1$ as
\begin{equation}
    e_1\rightarrow e_1\cdot N.
\end{equation}
Similarly, gauging the quantum electric 1-form symmetry $\mathbb{Z}^{(1)}_{N,e}$ has the following effect
\begin{equation}
    e_1\rightarrow e_1/N.
\end{equation}
From these relations one can check that gauging the symmetry and then its quantum dual gives back the original theory.

Gauging the 2-group in $\mathcal{T}_A$, along the lines of \autoref{sec.3}, amounts to gauging the quantum electric 1-form symmetry $\mathbb{Z}^{(1)}_{N,n,A}$ followed by the magnetic symmetry $\mathbb{Z}^{(0)}_{N,m,C}$. The resulting theory $\mathcal{T}_C$ has the same symmetry content as $\mathcal{T}_A$, but the roles played by the various magnetic and electric symmetries are exchanged:
\begin{equation}\begin{split}
    \mathcal{T}_A:\quad \mathbb{G}_{\hat{A},C}\times \mathbb{Z}_{N,m,B}^\text{non-inv},\\
    \mathcal{T}_C:\quad \mathbb{G}_{\hat{C},B}\times \mathbb{Z}_{N,m,A}^\text{non-inv}.
\end{split}\end{equation}
Hence, gauging gives back the same theory provided that the $U(1)_A$ coupling in $\mathcal{T}_A$, denoted $e_1^A$, equals the $U(1)_C$ coupling in $\mathcal{T}_C$, denoted $e_2^C$. Similarly for $e_1^B\rightarrow e_2^A$ and $e_1^C\rightarrow e_2^B$. Under the gauging of the electric 1-form symmetry in the first factor and the magnetic 0-form symmetry in the third factor, the second factor being unaffected, we find
\begin{equation}
    \mathcal{T}_A:\quad (e_1^A,e_1^B,e_1^C)\quad\longrightarrow\quad\mathcal{T}_C:\quad (e_1^A/N,e_1^B,e_1^C\cdot N)=(e_2^A,e_2^B,e_2^C).
\end{equation}
As discussed above, we need to impose
\begin{equation}
    e_1^A=e_2^C=e_1^C\cdot N,\quad e_1^B=e_2^A=e_1^A/N,\quad e_1^C=e_2^B=e_1^B,
\end{equation}
which admits the family of solutions
\begin{equation}
    (e_1^A,e_1^B,e_1^C)=(e_1^A,e_1^A/N,e_1^A/N),
\end{equation}
where $e_1^A$ is arbitrary.

\subsection{Product Theories}
The previous example belongs to a broader class of theories admitting non-invertible duality defects associated with half-spacetime gauging of a 2-group. Consider a theory $\mathcal{Q}$ with an anomaly-free $\mathbb{Z}_{N}^{(0)}$ symmetry. We then consider the product theory
\begin{equation}\label{eq.product}
    \mathcal{T}=\mathcal{Q}\times\mathcal{Q}\times\mathcal{Q},
\end{equation}
with anomaly theory as in \eqref{eq.4danomT}. Then, gauging the $\mathbb{Z}_N^{(0)}$ in the first $\mathcal{Q}$ factor gives a theory
\begin{equation}\label{eq.product1}
    \mathcal{T}_A=(\mathcal{Q}/\mathbb{Z}_N^{(0)})\times\mathcal{Q}^\text{non-inv}\times\mathcal{Q},
\end{equation}
with a 2-group between the symmetries of the first and third factor $(\mathcal{Q}/\mathbb{Z}_N^{(0)})\times\mathcal{Q}$, while the symmetry of the second factor becomes non-invertible. 

Gauging the 2-group in $\mathcal{T}_A$ yields the theory
\begin{equation}\label{eq.product2}
    \mathcal{T}_C=\mathcal{Q}^\text{non-inv}\times\mathcal{Q}\times(\mathcal{Q}/\mathbb{Z}_N^{(0)}),
\end{equation}
where the 2-group is now realized between the symmetries of last two factors. By the cyclic structure of the construction, the theories \eqref{eq.product1} and \eqref{eq.product2} are isomorphic,
\begin{equation}
    \mathcal{T}_A\cong\mathcal{T}_C.
\end{equation}
The $U(1)\times U(1)\times U(1)$ gauge theory of \autoref{sec.4.1} is a particular instance of this class of product theories. As noted in \cite{Choi:2024rjm}, one may also introduce interactions coupling the different sectors. An explicit instance is presented in the next subsection.

\subsection{Cyclic Quiver}
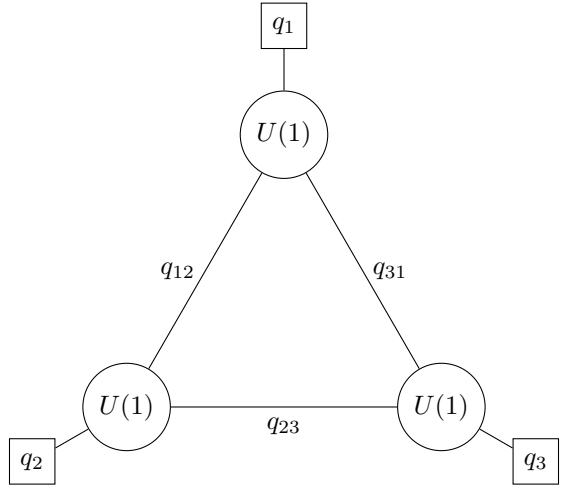
\begin{figure}[h!]\centering
    \begin{tikzpicture}[scale=1.2, every node/.style={font=\small}]
    \node[circle, draw, minimum size=10mm] (G1) at (90:2) {$U(1)$};
    \node[circle, draw, minimum size=10mm] (G2) at (210:2) {$U(1)$};
    \node[circle, draw, minimum size=10mm] (G3) at (330:2) {$U(1)$};
    \draw (G1) -- (G2) node[midway,left]{$q_{12}$};
    \draw (G2) -- (G3) node[midway,below]{$q_{23}$};
    \draw (G3) -- (G1) node[midway,right]{$q_{31}$};
    \node[draw, rectangle, minimum size=6mm] (F1) at (90:3.2) {$q_1$};
    \node[draw, rectangle, minimum size=6mm] (F2) at (210:3.2) {$q_2$};
    \node[draw, rectangle, minimum size=6mm] (F3) at (330:3.2) {$q_3$};
    \draw (G1) -- (F1);
    \draw (G2) -- (F2);
    \draw (G3) -- (F3);
    \end{tikzpicture}
    \caption{Quiver theory representing the moduli space of $PSU(3)$ instantons on $\mathbb{C}^2/\mathbb{Z}_3$. The Kronheimer-Nakajima quiver is found by setting $q_1=q_2=q_3=q_{12}=q_{13}=1$.}
    \label{fig.quiver}
\end{figure}

We consider the a generalization of the Kronheimer-Nakajima quiver \cite{PeterBKronheimer:1990zmj} shown in \autoref{fig.quiver}.
The associated 3d gauge theory preserves $\mathcal{N}=4$ supersymmetry. Each node represents a $U(1)$ vector multiplet, accompanied by a single hypermultiplet of charge $q_i$ transforming in the fundamental representation of $U(1)$. Each pair of nodes is connected by a single bifundamental hypermultiplet of charges $(q_{ij},-q_{ij})$. We assume that all matter-field charges are equal to a common integer $N$
\begin{equation}
    q_i=N,\qquad |q_{ij}|=N,\qquad\forall\,i,j=1,2,3,\quad i\neq j.
\end{equation}
In principle, it would suffice to assume that the greatest common divisor of the $q_i$ and $q_{ij}$ is $N$. With these choices, the quiver theory can be cast in the product form \eqref{eq.product}, with interactions among the factors.

This quiver theory is associated with the moduli space of $PSU(3)\cong U(3)/U(1)$ instantons on $\mathbb{C}^2/\mathbb{Z}_3$. The holonomy of the gauge field at infinity breaks $PSU(3)$ to $U(1)^3/U(1)$. Each $U(1)$ node is overbalanced as $N_f>2$, hence it is a ``good'' theory which flows, in the IR, to an interacting SCFT \cite{Gaiotto:2008ak}. The Coulomb 0-form symmetries of this theory include a $U(1)$ magnetic symmetry for each unbalanced gauge node \cite{Bhardwaj:2023zix,Nawata:2023rdx}
\begin{equation}
    \text{$0$-form symmetry:}\quad U(1)_A^{(0)}\times U(1)_B^{(0)}\times U(1)_C^{(0)}.
\end{equation}
The electric 1-form symmetry is determined by identifying the Wilson lines that are not screened by the matter fields. The fundamental hypermultiplets of charge $N$ at each node restrict the $U(1)_i^{(1)}$ symmetry to a $\mathbb{Z}_{N,i}^{(1)}$ subgroup. The bifundamentals further reduce these to a single diagonal $\mathbb{Z}_N^{(1)}$ subgroup,
\begin{equation}\label{eq.el1form}
    \text{$1$-form symmetry:}\quad \diag(\mathbb{Z}_{A,N}^{(1)}\times \mathbb{Z}_{B,N}^{(1)}\times \mathbb{Z}_{C,N}^{(1)})\cong \mathbb{Z}_{N}^{(1)}.
\end{equation}
In addition, the theory possesses 0-form flavour symmetries, which however play no role in what follows.

We stack this theory with an anomaly theory of the form \eqref{eq.4danomT} for the subgroup $\mathbb{Z}_{N,A}^{(0)}\times\mathbb{Z}_{N,B}^{(0)}\times\mathbb{Z}_{N,C}^{(0)}\subset U(1)_A^{(0)}\times U(1)_B^{(0)}\times U(1)_C^{(0)}$. Gauging $\mathbb{Z}_{N,A}^{(0)}$ then affects the gauge coupling $e_A$ of the first node and, in addition, generates a 2-group structure between the quantum symmetry $\hat{\mathbb{Z}}_{N,A}^{(1)}$ and the magnetic 0-form symmetry $\mathbb{Z}_{N,C}^{(0)}$. The symmetry $\hat{\mathbb{Z}}_{N,A}^{(1)}$ is identified with the electric 1-form symmetry \eqref{eq.el1form}. Moreover, the symmetry $\mathbb{Z}_{N,B}^{(0)}$ becomes non-invertible. We are then in a position to gauge the 2-group. As in the previous examples, the theory is invariant under half-spacetime gauging of the 2-group. This invariance follows from the symmetries of the quiver diagram, which form the dihedral group $D_3$ of an equilateral triangle. The duality defect realizes the cyclic subgroup $\mathbb{Z}_3\subset D_3$.

\section{Conclusions}\label{sec.5}
In this article we have constructed a class of non-invertible duality defects in three-dimensional quantum field theories arising from half-spacetime gauging of a 2-group symmetry. To the best of our knowledge, this provides the first explicit example of a non-invertible duality defect produced by gauging a 2-group structure.

Our starting point was a parent theory $\mathcal{T}$ equipped with two discrete and Abelian 0-form symmetries with a prescribed mixed anomaly \cite{Kapustin:2013uxa, Benini:2018reh}. Gauging one of the two factors yields a theory with a 2-group symmetry, while gauging the other produces a theory with a non-invertible 0-form symmetry, whose fusion rules we derived explicitly in \eqref{enoninvrules}-\eqref{enoninvrules2}. Although the existence of the 2-group structure in the first case was already known in the literature, the non-invertible structure in the latter case, to the best of our knowledge, had not been previously analyzed.

Building on these results, in \autoref{sec.3} we considered a parent theory with three 0-form symmetries and a cyclic anomaly structure, in which gauging different factors leads to mutually dual theories. Under this assumption, the half-spacetime gauging of the 2-group becomes a genuine symmetry, implemented by a duality defect $\mathcal{D}$ whose fusion rules we derived in \eqref{efusiondualitydefect}. We illustrated this construction with explicit examples in \autoref{sec.4}, including a $U(1)\times U(1)\times U(1)$ gauge theory, a general class of product theories and a generalization of the cyclic Kronheimer-Nakajima quiver of \cite{PeterBKronheimer:1990zmj}.

\paragraph{Future Directions.}
\begin{itemize}
    \item A first natural extension concerns more generic higher-categorical structures. Our construction relies on a 2-group with trivial group homomorphism $\rho$. In our setup this arises naturally as we started from the parent theory $\mathcal{T}$ with anomaly theory \eqref{eq.4danomT}. Starting instead from the theory with the 2-group, it would be possible to include non-trivial $\rho$. It would be interesting to understand how the fusion rules of both non-invertible symmetries are affected. More generally, one can consider higher $n$-groups, and study the structure of the resulting duality defects.
    \item A second avenue concerns the SymTFT description of the duality defects constructed here. In the SymTFT framework \cite{Gaiotto:2020iye,Apruzzi:2021nmk,Burbano:2021loy,Freed:2022qnc,Kaidi:2022cpf}, non-invertible symmetries and their fusion rules are encoded by a choice of topological boundary conditions of a bulk topological field theory in one dimension higher. It would be interesting to identify the bulk theory associated with the 2-group anomaly considered in this paper and to recover the duality defect $\mathcal{D}$, together with its fusion rules \eqref{efusiondualitydefect}, as a topological interface in the bulk. 
    \item A third direction concerns concrete physical realizations. The examples discussed in \autoref{sec.4} are based on simple gauge-theoretic and quiver constructions while many further realizations are likely to exist. In particular, supersymmetric theories with discrete higher-form symmetries, possibly engineered from string theory or from the geometry of internal manifolds \cite{Albertini:2020mdx, Apruzzi:2021phx, Bhardwaj:2023kri}, provide a natural arena where 2-group gauging and the associated duality defects could be systematically explored. The most natural starting point would be to consider a generalization of the compactification of a 6d $\mathcal{N}=(2,0)$ theory of type $A_{N-1}$ \cite{Gukov:2020btk,Bashmakov:2022jtl,Chen:2022vvd,Bashmakov:2022uek,Bashmakov:2023kwo,Chen:2023qnv,Gukov:2025dol}, to 6d little string theories. Indeed, as shown in \cite{DelZotto:2020sop}, these theories possess a 2-group symmetry. Then, compactifying on a three-manifold, would give rise to a large class of 3d theories with a 2-group symmetry. Duality defects obtained by gauging the 2-group are related to the mapping class group of the three-manifold.
\end{itemize}

\section*{Acknowledgment}
The authors are grateful to Babak Haghighat for useful discussions. DB is especially thankful to Anant Shri for his continuous support throughout the elaboration of this article. WC is supported by National Natural Science Foundation of China under Grant No. 42530108. LR acknowledges partial support by the INFN.

\appendix

\section{Conventions}\label{aconventions}
\begin{enumerate}
    \item Lorentzian physical spacetime of dimension $D$ is denoted $X_D$, with $k$-dimensional submanifolds written $X_k$ (possibly with additional labels). We denote with $Y_{D+1}$ the auxiliary space encoding the anomaly, with an analogue notation for its submanifolds.
    \item $C_k=C_k(M,R)$ denotes the $k$-chains over a ring $R$ for a given triangulation of a manifold $M$; we usually omit $M$ and take $R=\mathbb{Z}$. The $k$-cycles and $k$-boundaries are denoted $Z_k$ and $B_k$ (closed and exact chains, respectively) and $H_k=Z_k/B_k$ is the $k$-th homology group.
    \item $C^k=C^k(M,G)$ denotes the $k$-cochains valued in an Abelian group $G$. We write $C^k_A$ or $C^k_{\hat{A}}$ respectively when $G=G_A$ or $G=\hat{G}_A=\mathrm{Hom}(G,\mathbb{R}/\mathbb{Z})$ (the Pontryagin dual). The cocycles and coboundaries are $Z^k$ and $B^k$ (closed and exact cochains) and $H^k=Z^k/B^k$ is the $k$-th cohomology group. See \appref{aintrocohomology} for an introduction to chains and cochains.
    \item $G_A^{(q)}$ denotes a $q$-form group symmetry, with defect network specified by a cocycle $A^{(q+1)}$. The Pontryagin dual symmetry is $\hat{G}_A^{(D-q-2)}$, with associated cocycle $\hat{A}^{(D-q-1)}$. Lower-case letters ($a^{(q+1)}$, $\hat{a}^{(D-q-1)}$) correspond to a gauged background.
    \item Writing $S_\text{anomaly}[A,\phi]=\int_{Y_{D+1}}\mathcal{L}_\text{anomaly}[A,\phi]$, we define
    \begin{align}\label{eDeltaSan}
        \Delta S_\text{anomaly}[A,\lambda,\phi]= S_\text{anomaly}[A+\delta\lambda,\phi]- S_\text{anomaly}[A,\phi]=\int_{X_D} \Delta\mathcal{L}_\text{anomaly}[A,\lambda,\phi]\,,
    \end{align}
    so that
    \begin{align}\label{eanomalytheoryconv}
        Z[A+\delta\lambda]=\int\mathcal{D}\phi\ e^{i S[A,\phi]+i\Delta S_\text{anomaly}[A,\lambda,\phi]}D_{A+\delta\lambda}\,,
    \end{align}
    where $D_{A+\delta\lambda}$ is the defect network specified by $A+\delta\lambda$. The same convention extends to higher-group symmetries.
    \item For a compact space $X$ which is split along a codimension-1 surface into two halves, $a^{k}_{\geq 0}$ denotes a $k$-cochain in the (absolute) cohomology of $X_{\geq 0}$ and $a^{k}_{>0}$ the relative cochain that vanishes on the boundary. The same notation applies to further subdivisions of $X$, see \appref{ahalfspacetimegauging}.
    \item We adopt the following shorthand throughout. The symbol $\sum_{a\in H^q}$ denotes a sum over a choice of independent representative cocycles in $Z^q$. Cup products, including Steenrod cup products, are usually suppressed in the notation. The pairing of a cochain $\alpha$ with the fundamental chain of $M$ is written $\int_M\alpha$ (see \appref{acochainintegration}). Moreover, with our orientation conventions, $\int_M \delta\beta=\int_{\partial M}\beta$. Finally, we refer to a change of cocycle representative as a \textit{gauge transformation} and to invariance under such a change as \textit{gauge invariance}.
\end{enumerate}

\section{Introduction to Chains and Cochains}\label{aintrocohomology}
We give a brief, physics-oriented review of chains, cochains and related tools, summarising the results used in the main text. We mainly follow \cite{fatibene2021relativistic,Benini:2018reh}. For a more axiomatic treatment we recommend \cite{Hatcher2002}.

\subsection{Chains (and Connections to Manifold Triangulations)}\label{achain}
\begin{figure}[h]
    \begin{center}
    \includegraphics[scale=0.8]{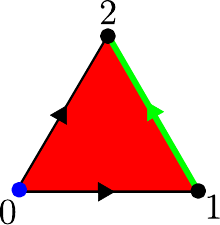}
    \caption{In this figure we depict the 0-simplex $[0]$ in blue, the 1-simplex $[1,2]$ in green and the 2-simplex $[0,1,2]$ in red. The orientation runs from smaller to larger indices.}\label{f2simplex}
    \end{center}
\end{figure}

Take a finite collection of distinct points $x_i\in\mathbb{R}$, forming the \textit{set of vertices} $V=\{x_i\}$. An ordered tuple $[x_{i_0},\,\hdots,\, x_{i_k}]$ with $x_{i_a}<x_{i_{a+1}}$ is called an \textit{(oriented abstract) $\mathit{k}$-simplex}. We then define a $(k-1)$\textit{-face} of $[x_{i_0},\,\hdots,\,x_{i_k}]$ as $[x_{i_0},\,\hdots,\,\hat{x}_{i_j},\,\hdots,\,x_{i_k}]$, where the hat denotes a missing element (we can generalize this for more missing elements). The terminology is motivated geometrically: each $x_i$ is a vertex and $[x_{i_0},\,\hdots,\, x_{i_k}]$ represents an \textit{elementary $k$-dimensional cell} with its corresponding $(k-1)$-dimensional faces (see \autoref{f2simplex}). The collection of all $k$-simplices and their faces is called a \textit{simplicial complex} $\Delta$.
Given a manifold, one can construct a simplicial complex $\Delta$ from a suitable open cover and its intersections; we call this a \textit{(Poincaré dual) triangulation}, see \cite{fatibene2021relativistic,Hatcher2002} for more explicit constructions.
 
Given $\Delta$, for each $k$ we consider the set of $k$-simplices $\{[x_{i_0},\,\hdots,\, x_{i_k}]\}$ and define an $R$-module $C_k=C_k(\Delta,R)$ over a ring $R$. Elements of $C_k$ are called \textit{$\mathit{k}$-chains} and, for the purpose of this article, we fix $R=\mathbb{Z}$. Concretely, $C_k$ consists of $\mathbb{Z}$-linear combinations of the $k$-simplices, each simplex playing the role of an independent generator. Geometrically, a $k$-chain represents an embedded (possibly disconnected) $k$-dimensional surface, with simplices counted with multiplicity and orientation. This becomes relevant once we introduce cochains and the corresponding notion of integration over chains. 
 
We then define the \textit{boundary operator} $\partial:C_k\rightarrow C_{k-1}$ in the following way on all $k$-simplexes
\begin{align}
    \partial[x_{i_0},\,\hdots,\, x_{i_k}]=\underset{j=0}{\overset{k}{\sum}}(-1)^j[x_{i_0},\,\hdots,\,\hat{x}_{i_j},\,\hdots,\,x_{i_k}]\,,
\end{align}
and we extend the action on the full $C_k$ by $\mathbb{Z}$-linearity. As the name suggests, this defines a new chain, which we identify with the boundary of the $k$-simplex. It is possible to prove that $\partial^2=0$. 
 
\begin{figure}[h]
    \begin{center}
    \includegraphics[scale=0.8]{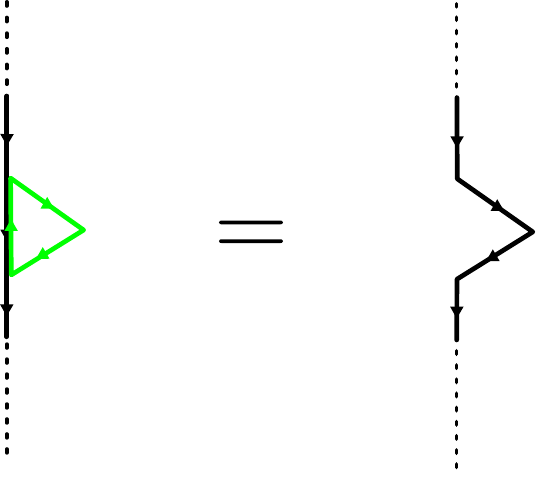}
    \caption{In this figure we depict some 1-chain in black, that might wrap a non-trivial 1-cycle, and in green a 1-boundary. The 1-boundary can be shrunk on the triangle it encircles. Moreover, we notice that summing this 1-boundary to the 1-chain applies something we can interpret as a topological move.}\label{fig.faddingboundary}
    \end{center}
\end{figure}

We define the set of $k$-\textit{cycles} $Z_k=\{\sigma\in C_k|\partial\sigma=0\}$ (these are called cycles precisely because they have no boundary). We then define $k$-\textit{boundaries} as $B_k=\{b\in C_k|\exists\sigma\in C_{k+1}\rightarrow \partial\sigma=b\}$. These are sometimes called \textit{contractible cycles}, since the higher-dimensional chain $\sigma$ with $\partial\sigma=b$ provides an explicit contraction of $b$ to a point (see \autoref{fig.faddingboundary}). Adding a $k$-boundary to a $k$-chain therefore amounts to a topological move (again, refer to \autoref{fig.faddingboundary}).
Finally we define the $k^{\text{th}}$\textit{-homology group} $H_k=\tfrac{Z_k}{B_k}$, since both $Z_k$ and $B_k$ are Abelian groups under $\mathbb{Z}$-linear addition. Its generators correspond to inequivalent non-contractible cycles.

\subsection{Cochains (and Connections to Group Symmetry Defects)}\label{acochain}
\begin{figure}[h]
    \begin{center}
    \includegraphics[scale=0.8]{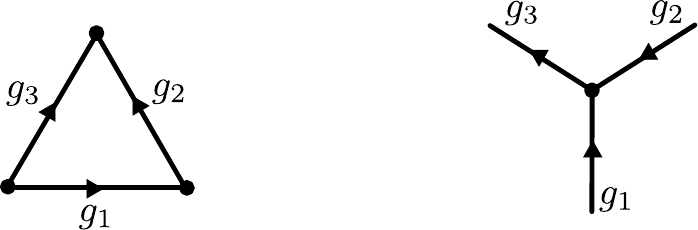}
    \caption{On the left we depict a $1$-cochain as an assignment of group elements on a simplex. On the left we depict the same in the Poincaré dual picture.}\label{fig.fcochains}
    \end{center}
\end{figure}

For an Abelian group $G$, the set of \textit{$\mathit{k}$-cochains} is denoted $C^k=C^k(\Delta,G)$. These are $\mathbb{Z}$-linear maps $\alpha:C_k\rightarrow G$, namely
\begin{align}
    \alpha(k_1\sigma_1+k_2\sigma_2)=\alpha(\sigma_1)^{k_1}\cdot_{G}\alpha(\sigma_2)^{k_2}\,,
\end{align}
$\forall k_i\in\mathbb{Z},\,\forall\sigma_i\in C_k$, where $\cdot_G$ denotes multiplication in $G$ and $\alpha(\sigma)^k$ stands for the $k$-fold product of $\alpha(\sigma)$. Similarly we can also define a composition of cochains in the following way
\begin{align}\label{ecompositioncochains}
    (k_1\alpha+k_2\beta)(\sigma)=\alpha(\sigma)^{k_1}\cdot_G\beta(\sigma)^{k_2}\,.
\end{align}
Concretely, specifying a cochain amounts to assigning a group element to each generating simplex of $C_k$, namely we have to find a collection of the following elements
\begin{align}\label{ecochainbase}
\alpha\left([x_{i_0}\hdots x_{i_k}]\right)=\alpha_{x_{i_0}\hdots x_{i_k}}\in G\,.
\end{align}
This corresponds to assigning a group element to each $k$-simplex in a triangulation, see \autoref{fig.fcochains}. Such assignments correspond to networks of intersecting defects implementing group transformations, as becomes apparent in the Poincaré dual picture (see \autoref{fig.fcochains}).

\noindent
We then define the \textit{coboundary operator} $\delta:C^k\rightarrow C^{k+1}$ as follows
\begin{align}\label{ecoboundary}
\delta\alpha(\sigma)\equiv\alpha(\partial\sigma)\,,
\end{align}
$\forall \alpha \in C^k,\,\forall\sigma\in C_{k+1}$. One verifies that $\delta$ is linear with respect to the cochain addition \eqref{ecompositioncochains}.
 
Given that $\partial^2=0$, one can immediately prove that $\delta^2=0$. We can then define the set of \textit{cocycles} $Z^k=\{\forall\alpha\in C^k|\delta\alpha=0\}$, where $\delta\alpha=0$ means that $\delta\alpha$ sends every $\sigma\in C_{k+1}$ to the identity element of $G$. This is equivalent to guaranteeing that the oriented product of all group elements associated to the boundary of any simplex equals the group identity (e.g. in \autoref{fig.fcochains} this means $g_3=g_1 g_2$). Equivalently, in the Poincaré dual picture, the cocycle condition ensures that the defect network is consistent with the group multiplication law, which is precisely why cocycles are the natural objects for describing such networks.
 
By an argument analogous to the one illustrated in \autoref{fig.faddingboundary} for chains, coboundaries correspond to topological moves of the defect network.
 
We can define the set of the coboundaries $B^k=\{\forall\alpha\in C^k|\exists\beta\in C^{k-1}\rightarrow \alpha=\delta\beta\}$ and finally the $k^{\text{th}}$\textit{-cohomology group} $H^k=\tfrac{Z^k}{B^k}$. Cohomology classes thus parametrize all topologically inequivalent defect networks.

\subsection{Eilenberg–MacLane Spaces and Homotopy Interpretation of Cochains}\label{aeilenbermaclane}
For this subsection we refer mostly to \cite{Hatcher2002, Tachikawa:2017gyf}, giving a brief account on Eilenberg-MacLane spaces and their use in the context of this article. 

Let $K(G,k)=B^k G$ denote (any representative of) the homotopy class of spaces whose only non-trivial homotopy group is $\pi_k=G$. The \textit{classifying space} is a special case $K(G,1)=B^1 G=BG$. For example, $B\mathbb{Z}=K(\mathbb{Z},1)$ admits $S^1$, $S^1\times\mathbb{R}$, \dots\ as representatives. Their use in this context is due to the fact that there is a bijection between $H^k(M,G)$, which we are interested in, and the set of homotopy classes of maps between the reference manifold $M$ and $K(G,k)$.\footnote{This is an equivalence class between maps among two manifold $X$ and $Y$. We write $f\sim g$ if, for $f,g:X\rightarrow Y$, there exists a continuous $F:X\times[0,1]\rightarrow Y$ such that $F(x,0)=f(x)$ and $F(x,1)=g(x)$, $\forall x\in X$.} Due to this bijection, in the literature cochains are often identified with homotopy classes of maps \cite{Tachikawa:2017gyf, Benini:2018reh}, as we do as well in \eqref{eq.parentanomaly}.

\subsection{(Steenrod) Cup Product}\label{acup}
Given a $(p+q)$-simplex $\sigma=[x_{i_0},\,\hdots,\, x_{i_{p+q}}]$ we can define the \textit{front $p$-face} as $^p\sigma=[x_{i_0},\,\hdots,\, x_{i_p}]$ and the \textit{back $q$-face} as $\sigma^q=[x_{i_p},\,\hdots,\, x_{i_{p+q}}]$. Both definitions extend by $\mathbb{Z}$-linearity to all of $C_{p+q}(\Delta,\mathbb{Z})$. With this we define the \textit{cup product}
\begin{align}\label{ecupproduct}
    \left(\alpha\cup\beta\right)\left(\sigma\right)=\alpha(^p\sigma)\cdot_G\beta(\sigma^q)\,,
\end{align}
$\forall \alpha\in C^p(\Delta,G),\,\forall\beta\in C^q(\Delta,G),\,\forall\sigma\in C_{p+q}(\Delta,\mathbb{Z})$.
 
Given a map $\phi:G\times G'\rightarrow G''$, the \textit{Steenrod cup product} is defined by \cite{csteenrod, Benini:2018reh}
\begin{align}\label{estenroodscup}
    \begin{split}
    &\left(\alpha\cup_\phi\beta\right)(\sigma)=\phi\left(\alpha( ^p\sigma),\beta(\sigma^q)\right)\,,\\
    &\left(\beta\cup_\phi\alpha\right)(\sigma)=\phi\left(\alpha( \sigma^p),\beta(^q\sigma)\right)\,,
    \end{split}
\end{align}
$\forall \alpha\in C^p(\Delta,G),\,\forall\beta\in C^q(\Delta,G'),\,\forall\sigma\in C_{p+q}(\Delta,\mathbb{Z})$, where now the result is an element of $G''$. This generalizes the ordinary cup product, recovered when $G=G'=G''$ and $\phi$ is the group multiplication.
 
Consider the Pontryagin dual group $\hat{G}=\text{Hom}(G,\tfrac{\mathbb{R}}{2\pi\mathbb{Z}})$, where we denote the action of $\hat{g}\in \hat{G}$ on $g\in G$ by $\hat{g}(g)\in \tfrac{\mathbb{R}}{2\pi\mathbb{Z}}$. Then we can choose the following pairing $\Phi:G\times \hat{G}\rightarrow \tfrac{\mathbb{R}}{2\pi\mathbb{Z}}$
\begin{align}\label{eq.esteenrodpontr}
    \Phi(g,\hat{g})=\hat{g}(g)\,,
\end{align}
$\forall g\in G,\,\forall\hat{g}\in \hat{G}$. The Steenrod cup product with this pairing is the natural ingredient appearing, implicitly or explicitly, in the construction of finite-group anomalies, since one needs to build $\mathbb{R}/2\pi\mathbb{Z}$-valued phase factors out of the cocycles representing the defect networks. We return to this point in \appref{acochainintegration}.

\paragraph{Cup Product Properties.} 
Let $\alpha^{(p)}\in C^p$ and let $\beta^{(q)}\in C^q$. From \eqref{ecoboundary}-\eqref{ecupproduct}, one derives the graded Leibniz rule for the (Steenrod) cup product with respect to the coboundary operator:
\begin{equation}\label{egradedLeibnitz}
    \delta(\alpha^{(p)}\cup_\phi\beta^{(q)})=\delta\alpha^{(p)}\cup_\phi\beta^{(q)}+(-1)^p\alpha^{(p)}\cup_\phi\delta\beta^{(q)}\,.
\end{equation}
It is furthermore possible to prove that the $\cup_\phi$ is graded-commutative for cocycles, modulo exact cochains \cite{Benini:2018reh}
\begin{align}\label{egradedcommutativity}
    \alpha^{(p)}\cup_\phi\beta^{(q)}=(-1)^{pq}\beta^{(q)}\cup_\phi\alpha^{(p)}+\delta(\hdots)\,.
\end{align}

\subsection{Integration of Cochains}\label{acochainintegration}
In the literature one often finds  $\mathit{\alpha(\sigma)=\int_\sigma \alpha}$. There are multiple reasons for this.
 
One of the main reasons is that $\mathbb{Z}_N$ cochains can be represented by gauge fields with fixed holonomies, e.g. in DW-theories, then we do really have an integral. Moreover, when $G$ is a subset or some quotient of $\mathbb{C}$, so that then $\alpha(\sigma)$ is a number, then $\alpha(\sigma)$ is linear in the subspaces in which we can divide $\sigma$ (e.g. if $\sigma_1+\sigma_2=\sigma$, $\int_\sigma\alpha=\int_{\sigma_1}\alpha+\int_{\sigma_2}\alpha$). Additionally, this operation always respects a form of Stokes' theorem, since $\int_\sigma\delta\alpha=\int_{\partial\sigma}\alpha$ and hence integration by parts.
 
Integration over a manifold is defined by triangulating it and taking, as the integration chain, the sum of all top-dimensional simplices in the triangulation.
 
Such integrals appear naturally in the formulation of anomalies for finite groups. In a typical computation, one begins with a collection of cochains $\{A^{(q+1)}\}$ related to a defect network and we want to construct a phase factor. This is often achieved by the Steenrod cup product with the mapping chosen in \eqref{eq.esteenrodpontr}, although other operations on $\{A^{(q+1)}\}$ usually appear before this last operation (see e.g. \eqref{eq.parentanomaly}).
 
We are primarily interested in integrals over cycles, which is why homology and cohomology play a central role here. Indeed, if $\sigma=\partial\hat{\sigma}$ is a boundary and $\alpha$ is closed, $\int_\sigma\alpha=\int_{\hat\sigma}\delta\alpha=0$. Hence if we deform the integration by contractible cycles the result does not change (see \autoref{fig.faddingboundary}). The reason for passing to cohomology classes, rather than working with cocycles in $Z^k$ alone, is precisely that we integrate over cycles, namely on $\sigma|\partial\sigma=0$. On these elements the integral of an exact cochain vanishes $\int_\sigma\delta\alpha=\int_{\partial\sigma}\alpha=0$.

\subsection{Useful Cochain Identities}\label{app.eresfourpoinc}
This subsection summarizes some key results used throughout the article; readers unfamiliar with chains and cochains are referred to the first part of \appref{aintrocohomology}. Throughout, the subscripts $A$ and $\hat{A}$ on $k$-cochains $\phi^{(k)}$ (and on related sets) indicate, respectively, that the cochain takes values in a finite Abelian group $G_A$ or in its Pontryagin dual $\hat{G}_A$. We similarly use $\geq 0$ and $>0$ for the relative cochains, as explained in \appref{aconventions}.

We define the \textit{Euler counterterm} as
\begin{align}\label{eeulercounterterm}
    \chi_A=\frac{|H_A^0||H_A^2|\hdots}{|H_A^1||H_A^3|\hdots}=\frac{|C_A^0||C_A^2|\hdots}{|C_A^1||C_A^3|\hdots}\,.
\end{align}
The equality above expresses the well-known fact that the Euler characteristic is independent of the chosen triangulation. It can be proven using the following properties
\begin{align}\label{esomerelations}
    H^k_A=\frac{Z_A^k}{B_A^k},\ \ \ B_A^k\sim \frac{C_A^{k-1}}{Z_A^{k-1}},\ \ \ C^D_A=Z^D_A,\ \ \ Z^0_A=H^0_A\,.
\end{align}
where $D$ is the dimension of the triangulated manifold.

We finally mention some known properties in homology and cohomology theory, used extensively in the literature (see e.g. \cite{Kaidi:2022cpf, Schafer-Nameki:2023jdn, Hatcher2002} and references therein). For a compact manifold without boundary, the following identities hold:
\begin{align}\label{efundamentalfourier}
    \begin{split}
    &\frac{1}{|C^k_A|}\sum_{\phi\in C^k_A}e^{i\int \phi^{(k)}\hat{\phi}^{(D-k)}}=\delta_{\hat{\phi},0}\,,\\
    &C_A^k\sim C_{\hat{A}}^{D-k}\,.
    \end{split}
\end{align}
Analogous identities hold on a space with boundary, relating absolute cochains to those relative to the boundary. We list them below
\begin{align}\label{efundamentalfourierrelative}
    \begin{split}
    &\frac{1}{|C^k_{A,\geq 0}|}\sum_{\phi\in C^k_{A,\geq 0}}e^{i\int \phi^{(k)}_{\geq 0}\hat{\phi}^{(D-k)}_{>0}}=\delta_{\hat{\phi}_{>0},0}\,,\ \ \ \frac{1}{|C^k_{A,>0}|}\sum_{\phi\in C^k_{A,\geq 0}}e^{i\int \phi^{(k)}_{> 0}\hat{\phi}^{(D-k)}_{\geq0}}=\delta_{\hat{\phi}_{\geq0},0}\,,\\
    &C_{A,\geq0}^k\sim C_{\hat{A},>0}^{D-k}\,,\\
    &|C_{A,\geq 0}^k|=|C_{A,> 0}^k||C_A^k(\partial X_{\geq 0})|\,.
    \end{split}
\end{align}

\subsection{Poincaré Duality} 
Finally, for cohomology representatives integrated over cycles, we have
\begin{align}\label{epoincareduality}
    e^{i\int_{X_D} A^{(q+1)}\hat{A}^{(D-q-1)}}=e^{i\int_{X^{\hat{A}}_{q+1}} A^{(q+1)}_{\hat{A}}}=e^{i\int_{X_{D-q-1}^A}\hat{A}_A^{(D-q-1)}}\,,
\end{align}
where $X^A_{D-q-1}$ (respectively $X^{\hat{A}}_{q+1}$) is the chain Poincaré-dual to the defect network defined by $A^{(q+1)}$ (respectively $\hat{A}^{(D-q-1)}$).\footnote{There is a sign factor in the last term of \eqref{epoincareduality} due to \eqref{egradedcommutativity}, which we omit here.} Finally $A^{(q+1)}_{\hat{A}}\in H^{q+1}(X^A_{D-q-1},\tfrac{\mathbb{R}}{2\pi \mathbb{Z}})$ and $\hat{A}_A^{(D-q-1)}\in H^{D-q-1}(X^{\hat{A}}_{q+1},\tfrac{\mathbb{R}}{2\pi \mathbb{Z}})$ represent the phase factors associated with the group transformations along the defect networks of $A^{(q+1)}$ and $\hat{A}^{(D-q-1)}$.
 
For concreteness, we now give an explicit construction of $X^{\hat{A}}_{q+1}$ and $A_{\hat{A}}^{(q+1)}$. Let the fundamental chain of $X_D$ be $\underset{\{i_a\}|x_a<x_{a+1}}{\sum}[x_{i_0}\hdots x_{i_D}]$ and let us define, in the notation of \eqref{ecochainbase}, the following
\begin{align}\label{eexplicitinducedcochain}
    A^{(q+1)}_{\hat{A},i_{D-q-1}\hdots i_D}=\sum_{i_0\hdots i_{D-q-2}|x_a< x_{a+1}}\hat{A}^{(D-q-1)}_{i_0\hdots i_{D-q-1}}\left(A^{(q+1)}_{i_{D-q-1}\hdots i_D}\right)\,,
\end{align}
where we have used the notion of action of the Pontryagin dual group on group elements (see e.g. \eqref{eq.esteenrodpontr}). We then define a sub-chain consisting of those simplices $[x_{i_{D-q-1}}\hdots x_{i_D}]$ for which $A^{(q+1)}_{\hat{A},i_{D-q-1}\hdots i_D}\neq\mathds{1}$. The fundamental chain of $X^{\hat{A}}_{q+1}$ is then the sum of these simplices and a representative of $A_{\hat{A}}^{(q+1)}$ is fully specified by \eqref{eexplicitinducedcochain}, since by construction $\int_{X^{\hat{A}}_{q+1}}A^{(q+1)}_{\hat{A}}=\int_{X_D} A^{(q+1)}\hat{A}^{(D-q-1)}$. Note that, by virtue of this identity, $A^{(q+1)}_{\hat{A}}\in Z_A^{q+1}(X^{\hat{A}}_{q+1},\tfrac{\mathbb{R}}{2\pi \mathbb{Z}})$.

\section{Gauging in Full and Half-Spacetime}
In this appendix we provide a pedagogical discussion of gauging in both full and half-spacetime, fixing the normalization conventions used in the main text.

\subsection{Gauging a Non-Anomalous Symmetry}\label{app.gaugingnonan}
In this subsection we spell out explicitly what we mean by gauging a symmetry. Although the procedure is standard, it is useful to fix the normalization conventions and to clarify the precise steps used elsewhere in this article.

Consider a theory $\mathcal{T}$ with a non-anomalous discrete $q$-form symmetry group $G^{(q)}$. We denote $A^{(q+1)}$ a cocycle representing a defect network inserted in the vacuum and we write the partition function of $\mathcal{T}$ as $Z[A^{(q+1)}]$. The fact that we require no anomaly means $Z[A^{(q+1)}]=Z[A^{(q+1)}+\delta\lambda^{(q)}_A]$.

By \textit{gauging} we mean the projection onto group singlets in the Hilbert space; at the level of the partition function, this is implemented by
\begin{align}
    Z_a^A=\frac{|H_A^{q-1}||H_A^{q-3}|\hdots}{|H_A^{q}||H_A^{q-2}|\hdots}\underset{a\in H^{q+1}_A}{\sum}Z[a^{(q+1)}]\,,
\end{align}
where the summation runs over a choice of representatives in the cohomology,\footnote{We adopt this normalization for convenience, following \cite{Kaidi:2022cpf}. It can be interpreted as factoring out residual gauge redefinitions.} and we denote a potential dependence on this choice by writing $Z_a^A$. Other consistent projections of the Hilbert space require discrete torsion, or equivalently stacking a TQFT before gauging. This achieves a projection within a eigenvector subspace of the gauged group.

Requiring a symmetry to become invisible implies that topological manipulations of the defect-network should be inconsequential. Concretely this amounts to asking $Z^A=Z_a^A=Z_{a+\delta\lambda}^A$, which holds here precisely because of the absence of an anomaly. For obvious reasons, we refer interchangeably to this change in representatives and invariance thereof, as gauge transformation and gauge invariance respectively.

More generally, one can twist the definition of gauging by a coupling constructed via the Steenrod cup product (see \appref{acup}), as follows
\begin{align}\label{eq.gaugingfirst}
    Z_a^A[\hat{A}]=\frac{|H_A^{q-1}||H_A^{q-3}|\hdots}{|H_A^{q}||H_A^{q-2}|\hdots}\underset{a\in H^{q+1}_A}{\sum}e^{i\int \hat{A}^{D-q-1}a^{(q+1)}}Z[a^{(q+1)}]\,,
\end{align}
where $\hat{A}^{D-q-1}\in C^{D-q-1}(X_D,\hat{G}^{(D-q-2)})$. Gauge invariance requires $\delta\hat{A}^{D-q-2}=0$, moreover we can consider $\hat{A}^{D-q-1}\in H^{D-q-1}(X_D,\hat{G}^{(D-q-2)})$, as $\hat{A}^{D-q-1}\rightarrow \hat{A}^{D-q-2}+\delta\lambda^{D-q-2}_{\hat{A}}$ has no effect. 

In the gauged theory, $\hat{A}^{(D-q-1)}$ specifies Pontryagin-dual cycles, briefly discussed and explicitly constructed in \appref{app.eresfourpoinc}, on which a defect network of $\hat{G}^{(D-q-2)}$ transformations is supported (see \autoref{fig.fcochains}). In the language of \eqref{epoincareduality}, this is nothing but the defect insertion in the gauged path integral \cite{Schafer-Nameki:2023jdn}
\begin{align}\label{edefectrepresentation}
    D_{\hat{A}}=e^{i\int_{X_{q+1}^{\hat{A}}}a_{\hat{A}}^{(q+1)}}=e^{ i\int_{X_D} \hat{A}^{(D-q-1)}a^{(q+1)}}\,.
\end{align}
While $D_{\hat{A}}$ is fundamentally an operator in the Hilbert space, it is represented as a numerical weight when the path integral is expressed as a summation over $a^{(q+1)}$, as seen in \eqref{eq.gaugingfirst}. This topological operator is the \textit{quantum symmetry} operator, which is the standard term for any new symmetry arising upon gauging.

It is possible to prove, using \eqref{eeulercounterterm}-\eqref{efundamentalfourier}, the following
\begin{align}\label{edoubleq.gaugingnonan}
    Z^{A,\hat{A}}[\hat{\hat{A}}]=\chi^{-1}_A Z[(-)^{Dq+D+q}\hat{\hat{A}}]\,.
\end{align}
Gauging the quantum symmetry once more, modulo a potential charge conjugation, recovers the original theory. The additional term in front, for the purposes of the partition function itself, is inessential, but becomes important as we discuss duality defects in \autoref{sec.3}. In the same way, we might write
\begin{align}\label{edefectrepresentation2}
    D_A = e^{(-)^{Dq+D+q} i\int_{X^A_{D-q-1}}\hat{a}^{(D-q-1)}}\,,
\end{align}
although the operator $D_A$ in the original theory $\mathcal{T}$ appears written in that way only when its partition function is written as $(\mathcal{T}/G_A)/\hat{G}_A$.  

For pedagogical purposes, we now show this explicitly in the case of a 0-form symmetry $G_A^{(0)}$ and its dual 1-form symmetry $\hat{G}_A^{(1)}$ in 3d, where some technical steps deserve attention. In this case the double gauging gives
\begin{align}\label{especificexample}
    Z^{A,\hat{A}}[\hat{\hat{A}}^{(1)}]=\frac{|H^0_{\hat{A}}|}{|H^0_A||H^1_{\hat{A}}|}\underset{a\in H^1_A,\hat{a}\in H^2_{\hat{A}}}{\sum}e^{i\int \hat{a}^{(2)}(a^{(1)}+\hat{\hat{A}}^{(1)})}Z[a^{(1)}]\,,
\end{align}
where we have used the graded commutativity of the Steenrod cup product in the cohomology \eqref{egradedcommutativity}. One might expect to perform the summation over $\hat{a}$ first, producing a Kronecker $\delta$ that would in turn fix the sum over $a$. This cannot be done directly, however, because the Kronecker $\delta$ in \eqref{efundamentalfourier} arises only after summing over \textit{all} cochains. We first replace $\underset{\hat{a}\in H_{\hat{A}}^2}{\sum}\rightarrow \frac{1}{|B^2_{\hat{A}}|}\underset{\hat{a}\in Z_{\hat{A}}^2}{\sum}$, which is allowed since summing additionally over all exact shifts of $\hat a^{(2)}$ does not affect the summand, and then add a Lagrange multiplier in the following way
\begin{align}\label{eusefulintermediatesteps}
    \begin{split}
    Z^{A,\hat{A}}[\hat{\hat{A}}]&=\frac{|H^0_{\hat{A}}|}{|H^0_A||H^1_{\hat{A}}||B^2_{\hat{A}}||C^0_A|}\underset{a\in H^1_A,\hat{a}\in C^2_{\hat{A}},\phi^0\in C^0_A}{\sum}e^{i\int \hat{a}^{(2)}(a^{(1)}+\hat{\hat{A}}^{(1)})+\delta\hat{a}^{(2)}\phi^{(0)}}Z[a]\\
    &=\frac{|H^0_{\hat{A}}|}{|H^0_A||H^1_{\hat{A}}||B^2_{\hat{A}}||C^0_A||B^1_A|}\underset{a\in Z^1_A,\hat{a}\in C^2_{\hat{A}},\phi^0\in C^0_A}{\sum}e^{i\int \hat{a}^{(2)}(a^{(1)}+\hat{\hat{A}}^{(1)}-\delta\phi^{(0)})}Z[a]\,.
    \end{split}
\end{align}
In the second line we have also traded $\underset{a\in H_{A}^1}{\sum}\rightarrow \frac{1}{|B^1_{A}|}\underset{a\in Z_A^1}{\sum}$, which makes some future steps easier. We can finally write
\begin{align}
    \begin{split}
    Z^{A,\hat{A}}[\hat{\hat{A}}]
    &=\frac{|H^0_{\hat{A}}||C^2_{\hat{A}}|}{|H^0_A||H^1_{\hat{A}}||B^2_{\hat{A}}||C^0_A||B^1_A|}\underset{a\in Z^1_A,\phi^0\in C^0_A}{\sum}\delta\left(a^{(1)}+\hat{\hat{A}}^{(1)}-\delta\phi^{(0)}\right)Z[a]\\
    &=\frac{|H^0_{\hat{A}}||C^2_{\hat{A}}|}{|H^0_A||H^1_{\hat{A}}||B^2_{\hat{A}}||C^0_A||B^1_A|}\underset{a\in Z^1_A,\phi^0\in C^0_A}{\sum}\delta\left(a^{(1)}+\hat{\hat{A}}^{(1)}\right)Z[a+\delta\phi^0]\\
    &=\frac{|H^0_{\hat{A}}||C^2_{\hat{A}}|}{|H^0_A||H^1_{\hat{A}}||B^2_{\hat{A}}||B^1_A|}\underset{a\in Z^1_A}{\sum}\delta\left(a^{(1)}+\hat{\hat{A}}^{(1)}\right)Z[a]\\
    &=\frac{|H^0_{\hat{A}}||C^2_{\hat{A}}|}{|H^0_A||H^1_{\hat{A}}||B^2_{\hat{A}}||B^1_A|}Z[-\hat{\hat{A}}]\,.
    \end{split}
\end{align}
In the first step we have integrated $\hat{a}$. In the second step we have shifted the sum of $a$ by an exact quantity, which was the reason for trading the sum of $a$ in the cohomology to one on all closed chains. In the third step we used $Z[a]=Z[a+\delta\phi]$, which makes the summand $\phi$-independent and produces a factor of $|C^0_A|$. Finally we have evaluated the $\delta$-function. Using \eqref{esomerelations}-\eqref{efundamentalfourier}, one can show that the prefactor is precisely the Euler counterterm $\chi_A^{-1}(X_3)$ defined in \eqref{eeulercounterterm}.

\subsection{Half-Spacetime Gauging and Duality Defects}\label{ahalfspacetimegauging}
Half-spacetime gauging requires the notion of relative cohomology \cite{Hatcher2002, Kaidi:2022cpf}. We do not provide a detailed account here, as this lies beyond the scope of the article. We limit ourselves to recalling the relation, in the relevant case, between relative and absolute cohomology (the latter introduced in \appref{acochain}).

Assume that there is a region in a compact manifold without a boundary $X$ where the manifold is diffeomorphic to $I\times Y$, with $I\subset\mathbb{R}$. We choose a point $0\in I$ and divide the manifold accordingly into two halves. We then gauge a symmetry on $X_{\geq 0}$, which has boundary $\partial X_{\geq 0}\sim Y$. This gauging requires summing over the relative cohomology of $X_{\geq 0}$, namely $a_{>0}^{(q+1)}\in H^{q+1}_{A,>0}=H^{q+1}(G_A^{(q)},X_{\geq 0}|\partial X_{\geq 0})$. In this case we can represent $a_{>0}^{(q+1)}$ as\footnote{This is relates to imposing Dirichlet boundary conditions, which render the boundary of the gauging topological when the gauged theory is dual to the ungauged one \cite{Gaiotto:2019xmp, Choi:2021kmx}.}
\begin{align}\label{edirichletrelative}
    \begin{split}
    &a^{(q+1)}_{>0}\in Z^{q+1}_A(X_{\geq 0},G_A^{(q)})\,,\\
    &a^{(q+1)}_{>0}\big|_{\partial X_{\geq 0}}=0\,,\\
    &\lambda_{a,>0}^{(q)}\big|_{\partial X_{\geq 0}}=0\text{ as }a^{(q+1)}_{>0}\rightarrow a^{(q+1)}_{>0}+\delta\lambda_{a,>0}^{(q)}\,.
    \end{split}
\end{align}
The gauging takes the form of
\begin{align}\label{ehalfgauging}
    Z_{\mathcal{D}_0}[A^{(q+1)}_{\leq0}, \hat{A}^{(D-q-1)}_{\geq0}]=\frac{|H_{A,>0}^{q-1}||H_{A,>0}^{q-3}|\hdots}{|H_{A,>0}^{q}||H_{A,>0}^{q-2}|\hdots}\underset{a\in H^{q+1}_{A,>}}{\sum}e^{i\int_{_{\geq 0}} \hat{A}^{D-q-1}_{\geq0}a^{(q+1)}_{> 0}}Z[A_{\leq0}^{(q+1)}+a^{(q+1)}_{>0}]\,,
\end{align}
where the cocycles $A_{\leq0}^{(q+1)}$ and $a^{(q+1)}_{>0}$ are added using \eqref{ecompositioncochains}, where each is set to $\mathds{1}$ outside its respective half of space. The sources are allowed to be non-vanishing at the boundary, so that they belong to the (absolute) cohomology of their respective half-spaces, as introduced in \appref{acochain}.

The defect $\mathcal{D}$ appearing as a subscript in \eqref{ehalfgauging} absorbs all generators of the $\hat{G}_A^{(D-q-2)}$ global symmetry in $X_{\geq 0}$, as we can see from their representation \eqref{edefectrepresentation} and the condition \eqref{edirichletrelative}. The analogous statement holds for $G_A^{(q)}$ on the other side of the wall,\footnote{This becomes clear upon representing the path integral of $Z$ in terms of $Z^A$ via \eqref{edoubleq.gaugingnonan}: $A^{(q+1)}_{\leq 0}$ then couples to a sum over $\hat{a}^{(D-q-1)}_{<0}$.} thus we can write
\begin{align}\label{edualityabs}
    \mathcal{D}_0 D_{\hat{A}}=D_A\mathcal{D}_0=\mathcal{D}_0\,.
\end{align}
$\mathcal{D}$ is in general not topological unless the gauged theory is isomorphic to the former one.

\section{Derivations and Computational Details}
In this appendix we collect a number of computations whose details would have cluttered the main text.

\subsection{Descent Identities of $\zeta^{(2)}$}
Starting from \eqref{eq.descendant}, which we reproduce below for convenience
\begin{align}\label{edescendantap}
    (A^{(1)}+\delta\lambda_A^{(0)})^*\beta-A^{(1)*}\beta=\delta\zeta^{(2)}(A^{(1)},\lambda_A^{(0)})\,,
\end{align}
we can find multiple identities among $\zeta$ with different functional dependencies. Recall that $\beta\in H^3(K^1 G_A^{(0)},\hat{G}_B^{(2)})$ and that $A^{(1)}$ is interpreted as a homotopy class of maps from spacetime to $K^1 G_A^{(0)}$ (see \appref{aeilenbermaclane}).

We work out one of the non-trivial identities explicitly. We can use \eqref{edescendantap} to write
\begin{align}\label{e3relations}
    \begin{split}
    (a^{(1)}+\delta\lambda_1^{(0)}+\delta\lambda_2^{(0)})^*\beta-(a^{(1)}+\delta\lambda_1^{(0)})^*\beta &=\delta\zeta^{(2)}(a^{(1)}+\delta\lambda_1^{(0)},\lambda_2^{(0)})\,,\\
    (a^{(1)}+\delta\lambda_1^{(0)})^*\beta-(a^{(1)})^*\beta &=\delta\zeta^{(2)}(a^{(1)},\lambda_1^{(0)})\,,\\
    (a^{(1)}+\delta\lambda_1^{(0)}+\delta\lambda_2^{(0)})^*\beta-(a^{(1)})^*\beta&=\delta\zeta^{(2)}(a^{(1)}+\delta\lambda_1^{(0)},\lambda_2^{(0)})\,.
    \end{split}
\end{align}
Summing the first two expressions and subtracting the last one we obtain
\begin{align}
    \delta\left(\zeta^{(2)}(a^{(1)}+\delta\lambda_1^{(0)},\lambda_2^{(0)})+\zeta^{(2)}(a^{(1)},\lambda_1^{(0)})-\zeta^{(2)}(a^{(1)},\lambda_1^{(0)}+\lambda_2^{(0)})\right)=0\,.
\end{align}
It follows that $\zeta^{(2)}(a^{(1)}+\delta\lambda_1^{(0)},\lambda_2^{(0)})+\zeta^{(2)}(a^{(1)},\lambda_1^{(0)})-\zeta^{(2)}(a^{(1)},\lambda_1^{(0)}+\lambda_2^{(0)})$ is closed. Since $K^1 G_A^{(0)}$ has no non-trivial $2$-cycles, this closed cochain is in fact exact and we can write \cite{Kapustin:2013uxa}
\begin{align}\label{edescendattokappa}
    \zeta^{(2)}(a^{(1)}+\delta\lambda_1^{(0)},\lambda_2^{(0)})+\zeta^{(2)}(a^{(1)},\lambda_1^{(0)})-\zeta^{(2)}(a^{(1)},\lambda_1^{(0)}+\lambda_2^{(0)})=\delta \kappa(a^{(1)},\lambda_1^{(0)},\lambda_2^{(0)})\,.
\end{align}
Different combinations of relations of the form \eqref{edescendantap}, analogous to those used in \eqref{e3relations} and combined with the absence of $2$-cycles in $K^1 G_A^{(0)}$, yield many further identities. We write the main ones which we use throughout the article, ignoring the exact terms
\begin{align}
    &\zeta^{(2)}(a^{(1)}+\delta\lambda_1^{(0)},\lambda_2^{(0)})+\zeta^{(2)}(a^{(1)},\lambda_1^{(0)})-\zeta^{(2)}(a^{(1)},\lambda_1^{(0)}+\lambda_2^{(0)})=\delta(\hdots)\,,\label{e3zetaidentity}\\
    &\zeta^{(2)}(a^{(1)},\lambda^{(0)})+\zeta^{(2)}(a^{(1)}+\delta\lambda^{(0)},-\lambda^{(0)})=\delta(\hdots)\,,\label{e2zetaidentity}\\
    &\zeta^{(2)}(a^{(1)},\lambda^{(0)}+c^{(0)})-\zeta^{(2)}(a^{(1)},\lambda^{(0)})=\delta(\hdots)\,,\ \ \ \forall c^{(0)}\in Z_A^0=H_A^0\,,\label{ezetaclosedidentity}\\
    &\zeta^{(2)}(0,c^{(0)})=\delta(\hdots)\,,\ \ \ \forall c^{(0)}\in Z_A^0=H_A^0\,.\label{esinglezeta}
\end{align}

\subsection{A Topological Theory with Controlled Gauge Variance}\label{agaugevarth}
This subsection establishes the topological nature, modulo gauge variance, of the partition function $\mathcal{Z}_{a,B}$ defined in equation \eqref{enoninvtheory} and used to construct the non-invertible defect $\mathcal{N}_B$ in \autoref{sec.2.2}. We rewrite $\mathcal{Z}_{a,B}$ as follows
\begin{align}
    \begin{split}\label{esomestepsTQFT}
    \mathcal{Z}_{a,B}&=\frac{1}{|C^1_{\hat{A}}(X_2^B)||H^0_A(X_2^B)|}\underset{\phi\in C_A^0(X_2^B),\,\hat{\eta}\in C^1_{\hat{A}}(X_2^B)}{\sum}e^{-i\int_{X_2^B}\zeta^{(2)}_B(a^{(1)},-\phi^{(0)})+\hat{\eta}^{(1)}\left(a^{(1)}-\delta\phi^{(0)}\right)}\\
    &=\frac{1}{|H^0_A(X_2^B)|}\underset{\phi\in C_A^0(X_2^B)}{\sum}\delta_{a,\delta\phi}e^{-i\int_{X_2^B}\zeta^{(2)}_B(\delta\phi^{(0)},-\phi^{(0)})}\\
    &=\frac{1}{|H^0_A(X_2^B)|}\underset{\phi\in C_A^0(X_2^B)}{\sum}\delta_{a,\delta\phi}e^{i\int_{X_2^B}\zeta^{(2)}_B(0,\phi^{(0)})}\,.
    \end{split}
\end{align}
In the first line we integrated $\hat{\eta}$, in the second one we used the Kronecker $\delta$, in the third we used \eqref{e2zetaidentity}. The properties of $\zeta$ translate to $\zeta_B$ under integration over $X_2^B$ thanks to the defining identity \eqref{einitialdefofzetaB}. Finally, either the result vanishes or $a^{(1)}=\delta a^{(0)}$ when $a^{(1)}$ is restricted to a cochain on $X_2^B$. In this case the Kronecker $\delta$ is satisfied by $\phi^{(0)}=a^{(0)}+c^{(0)},\ \forall c^{(0)}\in Z^0_A=H^0_A$ and thus we have
\begin{align}
    \mathcal{Z}_{a,B}=\frac{1}{|H^0_A(X_2^B)|}\sum_{c\in H^0_A}e^{i\int_{X_2^B}\zeta^{(2)}_B(0,a^{(0)}+c^{(0)})}=e^{i\int_{X_2^B}\zeta^{(2)}_B(0,a^{(0)})}\,,
\end{align}
where we have used \eqref{ezetaclosedidentity} (and again the fact that we can translate the property to $\zeta_B$ if it is integrated over $X_2^B$). In either case, the partition function equals an expression independent of the triangulation, modulo the gauge-variant term, as we wished to prove.

\subsection{Derivation of the Non-Invertible Fusion Rules}\label{afusionrules}
In this subsection we derive the fusion rules stated in \eqref{enoninvrules}. Since we treat the theory hosting these fusion rules as a gauging of a parent theory \eqref{egaugingnoninv}, some of the topological defects, or some of the ingredients used to define them, appear simply as phase factors (see the comments below \eqref{edefectrepresentation}). We reproduce below the definition of $\mathcal{N}_B$ from \eqref{enoninvtheory}-\eqref{eq.nnoninvertible}:
\begin{align}\label{ennoninvertibleappendix}
    \begin{split}
    \mathcal{N}_B&=\mathcal{Z}_{a,B}D_B(a^{(1)})\,,\\
    \mathcal{Z}_{a,B}&=\frac{1}{|C^1_{\hat{A}}(X_2^B)||H^0_A(X_2^B)|}\underset{\phi\in C_A^0(X_2^B),\,\hat{\eta}\in C^1_{\hat{A}}(X_2^B)}{\sum}e^{-i\int_{X_2^B}\zeta^{(2)}_B(a^{(1)},-\phi^{(0)})+\hat{\eta}^{(1)}\left(a^{(1)}-\delta\phi^{(0)}\right)}\,,
    \end{split}
\end{align}
where the insertion of $\mathcal{N}_B$ in \eqref{egaugingnoninv} is given by \eqref{etentativereprnb}-\eqref{ereprnb}. $D_B$ is a group-like defect for some $G_B^{(0)}$, now gauge variant, that therefore composes as \eqref{ecompositioncochains}
\begin{align}
D_{B_1}D_{B_2}=D_{B_1+B_2}\,.
\end{align}
We choose the defect to be unitary, so that $D_B^\dag=D_{-B}$, which immediately implies $\mathcal{N}_B^\dag=\mathcal{N}_{-B}$.

In a similar fashion, the 1-form symmetry $\hat{G}_A^{(1)}$-defect is represented in \eqref{egaugenoninv} as in \eqref{edefectrepresentation}, namely by
\begin{align}\label{eahatdefect}
    D_{\hat{A}}=e^{i\int \hat{A}^{(2)}a^{(1)}}\,.
\end{align}

\subsubsection{Fusion of $\mathcal{N}_{B_1}$ and $\mathcal{N}_{B_2}$}
We now consider the composition $\mathcal{N}_{B_1}\mathcal{N}_{B_2}$ of two such defects living on the same 2-cycle.\footnote{If they live on different 2-cycles, they simply commute and do not compose, as can be shown by an analogous argument.}
\begin{align}\label{enb12beginfusion}
    \begin{split}
    \mathcal{N}_{B_1}\mathcal{N}_{B_2}&=\frac{D_{B_1+B_2}}{|H^0_A(X_2^B)|^2}\underset{\phi_1,\phi_2\in C_A^0(X_2^B)}{\sum}\delta_{a,\delta\phi_1}\delta_{a,\delta\phi_2}e^{-i\int_{X_2^B}\zeta_{B_1}^{(2)}(a^{(1)},-\phi_1)+\zeta_{B_2}^{(2)}(a^{(1)},-\phi_2)}\\
    &=\frac{D_{B_1+B_2}}{|H^0_A(X_2^B)|^2}\underset{\phi_1\in C_A^0(X_2^B),\,c\in Z_A^0(X_2^B)}{\sum}\delta_{a,\delta\phi_1}e^{-i\int_{X_2^B}\zeta_{B_1}^{(2)}(a^{(1)},-\phi_1)+\zeta_{B_2}^{(2)}(a^{(1)},-\phi_1-c^{(0)})}\\
    &=\frac{D_{B_1+B_2}}{|H^0_A(X_2^B)|^2}\underset{\phi_1\in C_A^0(X_2^B),\,c\in Z_A^0(X_2^B)}{\sum}\delta_{a,\delta\phi_1}e^{-i\int_{X_2^B}\zeta_{B_1}^{(2)}(a^{(1)},-\phi_1)+\zeta_{B_2}^{(2)}(a^{(1)},-\phi_1)}\\
    &=\frac{D_{B_1+B_2}}{|H^0_A(X_2^B)|}\underset{\phi_1\in C_A^0(X_2^B)}{\sum}\delta_{a,\delta\phi_1}e^{-i\int_{X_2^B}\zeta_{B_1+B_2}^{(2)}(a^{(1)},-\phi_1)}\,.
    \end{split}
\end{align}
In the first step, we integrated out the $\hat{\eta}_i$ and composed the $D_{B_i}$. In the second step we used the fact that $a^{(1)}=\delta\phi_1^{(0)}=\delta\phi_2^{(0)}$, so $\phi_2^{(0)}-\phi_1^{(0)}$ must be a 0-cocycle $c^{(0)}\in Z^0_A=H^0_A$ (this restriction makes one of the two Kronecker $\delta$'s trivial). In the third step we use \eqref{esinglezeta} together with its translation to $\zeta_B^{(2)}$, as explained below \eqref{esomestepsTQFT}. In the final step we have used the definition of $\zeta^{(2)}_B$ from \eqref{einitialdefofzetaB}-\eqref{eexplicitinducedcochain}, together with the cochain composition rule \eqref{ecompositioncochains}.

Finally, re-exponentiating the constraint imposed by the Kronecker $\delta$, we obtain
\begin{align}
    \mathcal{N}_{B_1}\mathcal{N}_{B_2}=\mathcal{N}_{B_1+B_2}\,.
\end{align}
Although the composition appears group-like, the inverse does not always exist, so the structure is not actually that of a group. Indeed, the only candidate for the identity arises by setting $B=B_1=-B_2$. However, in this case, we would have
\begin{align}\label{enoninvsurprise}
\mathcal{N}_{B}\mathcal{N}_{-B}=\frac{\mathds{1}}{|H^0_A(X_2^B)|}\underset{\phi_1\in C_A^0(X_2^B)}{\sum}\delta_{a,\delta\phi_1}\neq \mathds{1}\,.
\end{align}
We rewrite this in terms of other symmetry operators. Following \eqref{eahatdefect}, we define $D_{\hat{a}_B}$ as the 1-form defect supported on a 1-cycle inside $X_2^B$ and obtain
\begin{align}\label{ereexpressingsumofdefects}
    \begin{split}
    \underset{\hat{a}_B\in H^1_{\hat{A}}(X_2^B)}{\sum}D_{\hat{a}_B}&=\underset{\hat{a}_B\in H^1_{\hat{A}}(X_2^B)}{\sum}e^{i\int_{X_2^B} \hat{a}_B^{(1)}a^{(1)}}\\
    &=\frac{1}{|B^1_{\hat{A}}(X_2^B)|}\underset{\hat{a}_B\in Z^1_{\hat{A}}(X_2^B)}{\sum}e^{i\int_{X_2^B} \hat{a}_B^{(1)}a^{(1)}}\\
    &=\frac{1}{|B^1_{\hat{A}}(X_2^B)||C^0_A(X_2^B)|}\underset{\hat{a}_B\in C^1_{\hat{A}}(X_2^B),\,\phi\in C^0_A(X_2^B)}{\sum}e^{i\int_{X_2^B} \hat{a}_B^{(1)}a^{(1)}-\delta\hat{a}_B^{(1)}\phi^{(0)}}\\
    &=|H^0_{\hat{A}}(X_2^B)|\chi^{-1}_A(X_2^B)\underset{\phi\in C^0_A(X_2^B)}{\sum}\delta_{a,\delta\phi}\,.
    \end{split}
\end{align}
In the second step we replace $\underset{\hat{a}_B\in H_{\hat{A}}^1}{\sum}\rightarrow \frac{1}{|B^1_{\hat{A}}|}\underset{\hat{a}_B\in Z_{\hat{A}}^1}{\sum}$, which is allowed since the additional sum over exact shifts of $\hat{a}_B^{(1)}$ does not affect the summand. We then introduce a Lagrange multiplier, allowing us to take $\hat{a}_B^{(1)}\in C^1_{\hat{A}}$. Finally, summing over $\hat{a}_B^{(1)}$ and using \eqref{eeulercounterterm}-\eqref{esomerelations} yields the result.

Re-expressing the sum over Kronecker $\delta$ in \eqref{enoninvsurprise} as the sum of defects in \eqref{ereexpressingsumofdefects}, we obtain \eqref{enoninvrules}.

\subsubsection{Fusion of $\mathcal{N}_{B}$ and $D_{\hat{A}}$}
Assume that $D_{\hat{A}}=e^{i\int \hat{A}^{(2)}a^{(1)}}$ wraps a 1-cycle inside $X_2^B$, i.e., on the same surface where some $\mathcal{N}_B$ lives. In this case, there exists some $\hat{A}_B^{(1)}\in Z^1_{\hat{A}}(X_2^B)$ such that
\begin{align}\label{e2to1}
    D_{\hat{A}}=e^{i\int \hat{A}^{(2)}a^{(1)}}=e^{i\int_{X_2^B}\hat{A}_B^{(1)}a^{(1)}}\,.
\end{align}
It is immediate that $\mathcal{N}_B D_{\hat{A}}=D_{\hat{A}}\mathcal{N}_B$, and \eqref{enoninvrules2} follows from the computation below
\begin{align*}
    \begin{split}
    \mathcal{N}_BD_{\hat{A}}=\frac{e^{i\int_{X_2^B}\hat{A}_B^{(1)}a^{(1)}}}{|C^1_{\hat{A}}(X_2^B)||H^0_A(X_2^B)|}\underset{\phi\in C_A^0(X_2^B),\,\hat{\eta}\in C^1_{\hat{A}}(X_2^B)}{\sum}&e^{-i\int_{X_2^B}\zeta^{(2)}_B(a^{(1)},-\phi^{(0)})+\hat{\eta}^{(1)}\left(a^{(1)}-\delta\phi^{(0)}\right)}\\
    =\frac{e^{i\int_{X_2^B}\hat{A}_B^{(1)}a^{(1)}}}{|C^1_{\hat{A}}(X_2^B)||H^0_A(X_2^B)|}\underset{\phi\in C_A^0(X_2^B),\,\hat{\eta}\in C^1_{\hat{A}}(X_2^B)}{\sum}&e^{-i\int_{X_2^B}\hat{A}_B^{(1)}(a^{(1)}-\delta\phi^{(0)})}\\
    &e^{-i\int_{X_2^B}\zeta^{(2)}_B(a^{(1)},-\phi^{(0)})+\hat{\eta}^{(1)}\left(a^{(1)}-\delta\phi^{(0)}\right)}\\
    =\frac{1}{|C^1_{\hat{A}}(X_2^B)||H^0_A(X_2^B)|}\underset{\phi\in C_A^0(X_2^B),\,\hat{\eta}\in C^1_{\hat{A}}(X_2^B)}{\sum}&e^{-i\int_{X_2^B}\zeta^{(2)}_B(a^{(1)},-\phi^{(0)})+\hat{\eta}^{(1)}\left(a^{(1)}-\delta\phi^{(0)}\right)}\\
    &\hspace{-7.1 cm}=\mathcal{N}_B\,.
    \end{split}
\end{align*}
In the second step, we shifted the summation variable $\hat{\eta}^{(1)}\rightarrow \hat{\eta}^{(1)}-\hat{A}_B^{(1)}$. In the third step, we canceled the $e^{i\int_{X_2^B}\hat{A}_B^{(1)}a^{(1)}}$ factors and used the fact that $e^{i\int_{X_2^B}\hat{A}^{(1)}_B\delta\phi^{(0)}}=e^{i\int_{X_2^B}\delta\hat{A}^{(1)}_B\phi^{(0)}}=1$, since $\hat{A}_B^{(1)}\in Z^1_{\hat{A}}(X_2^B)$.

\bibliographystyle{JHEP}
\bibliography{main}

\end{document}